\documentclass[prb,letterpaper,twocolumn,showpacs,floatfix,superscriptaddress,amsmath,amsfonts,amssymb,preprintnumbers,citeautoscript]{revtex4}
\pdfoutput=1
\usepackage[T1]{fontenc}\usepackage[latin1]{inputenc}
\usepackage{dcolumn,graphicx,color,hvfloat,booktabs,microtype,afterpage,ocg} \graphicspath{{Figures/}}
\usepackage[charter]{mathdesign}
\usepackage{sidecap}
\renewcommand{\figurename}{Fig.}
\renewcommand{\tablename}{Table}
\makeatletter\renewcommand{\fnum@figure}[1]{\figurename~\thefigure~(color online).}\makeatother
\makeatletter\renewcommand{\fnum@table}[1]{\tablename~\thetable.}\makeatother

\newcount\hh \newcount\mm
\hh=\time \divide\hh by 60
\mm=\hh \multiply\mm by 60 \mm=-\mm
\advance\mm by \time
\def\now{\number\hh:\ifnum\mm<10{}0\fi\number\mm}

\newcommand{\ToggleLayer}[2]{%
  \leavevmode%
  \pdfstartlink user {
    /Subtype /Link
    /Border [0 0 0]%
    /A <<
      /S/JavaScript
      /JS (
         var aOCGs = this.getOCGs();
         var AllChecked = 1;
         var AllIndex = 0;
         for(var i=0; aOCGs && i<aOCGs.length;i++)
         {
         if(aOCGs[i].name == "#1") {aOCGs[i].state = !aOCGs[i].state;}
         if(!aOCGs[i].state && aOCGs[i].name != "all") {AllChecked = 0;}
         if(aOCGs[i].name == "all") {aOCGs[i].state = false; AllIndex=i;}
         }
         if(AllChecked == 1) {aOCGs[AllIndex].state = true;}
      )
    >>
  }#2%
  \pdfendlink%
}

\usepackage[colorlinks,plainpages=false,linkcolor=blue,urlcolor=blue,citecolor=black,pdfpagemode=UseNone,pdfstartview=FitBH]{hyperref}
\newcommand{\BFA }{\mbox{BaFe$_2$As$_2$}}
\newcommand{\BFNA }{\mbox{BaFe$_{2-x}$Ni$_x$As$_2$}}
\newcommand{\BFNAour }{\mbox{BaFe$_{1.91}$Ni$_{0.09}$As$_2$}}
\newcommand{\BFCA }{\mbox{BaFe$_{2-x}$Co$_x$As$_2$}}
\newcommand{\BFCAour }{\mbox{BaFe$_{1.85}$Co$_{0.15}$As$_2$}}

\newcommand{\ybco }{\mbox{YBa$_2$Cu$_3$O$_{6+y}$}}
\newcommand{\BKFA }{\mbox{Ba$_{1-x}$K$_x$Fe$_2$As$_2$}}
\newcommand{\x }{\mbox{$\chi^{\prime\prime}(\mathbf{Q},\omega)$}}
\newcommand{\Er }{\mbox{$\omega_{\rm res}$}}
\newcommand{\Erodd }{\mbox{$\omega_{\rm res,odd}$}}
\newcommand{\Ereven }{\mbox{$\omega_{\rm res,even}$}}

\begin{document}


\title{Symmetry of spin excitation spectra in the tetragonal paramagnetic\\and superconducting phases of 122-ferropnictides}

\author{J.\,T.~Park}
\affiliation{Max-Planck-Institut für Festkörperforschung, Heisenbergstraße 1, 70569 Stuttgart, Germany}

\author{D.\,S.\,Inosov}\email[E-mail: ]{d.inosov@fkf.mpg.de}
\affiliation{Max-Planck-Institut für Festkörperforschung, Heisenbergstraße 1, 70569 Stuttgart, Germany}

\author{A.\,Yaresko}
\affiliation{Max-Planck-Institut für Festkörperforschung, Heisenbergstraße 1, 70569 Stuttgart, Germany}

\author{S.\,Graser}
\affiliation{\mbox{Center for Electronic Correlations and Magnetism, Institute of Physics, University of Augsburg, D-86135 Augsburg, Germany}}

\author{D.\,L.~Sun}
\affiliation{Max-Planck-Institut für Festkörperforschung, Heisenbergstraße 1, 70569 Stuttgart, Germany}

\author{Ph.\,Bourges}
\affiliation{Laboratoire L{\' e}on Brillouin, CEA-CNRS, CEA Saclay, 91191 Gif-sur-Yvette Cedex, France}

\author{Y.\,Sidis}
\affiliation{Laboratoire L{\' e}on Brillouin, CEA-CNRS, CEA Saclay, 91191 Gif-sur-Yvette Cedex, France}

\author{Yuan Li}
\affiliation{Max-Planck-Institut für Festkörperforschung, Heisenbergstraße 1, 70569 Stuttgart, Germany}

\author{J.-H.~Kim}
\affiliation{Max-Planck-Institut für Festkörperforschung, Heisenbergstraße 1, 70569 Stuttgart, Germany}

\author{D.\,Haug}
\affiliation{Max-Planck-Institut für Festkörperforschung, Heisenbergstraße 1, 70569 Stuttgart, Germany}

\author{A.\,Ivanov}
\affiliation{Institut Laue-Langevin, 156X, 38042 Grenoble cedex 9, France}

\author{K.\,Hradil}
\affiliation{\mbox{Forschungsneutronenquelle Heinz Maier-Leibnitz (FRM II), Technische Universität München, D-85747 Garching, Germany}}
\affiliation{Institut für Physikalische Chemie, Universität Göttingen, 37077 Göttingen, Germany}

\author{A.\,Schneidewind}
\affiliation{\mbox{Forschungsneutronenquelle Heinz Maier-Leibnitz (FRM II), Technische Universität München, D-85747 Garching, Germany}}
\affiliation{Gemeinsame Forschergruppe HZB~--~TU Dresden, Helmholtz-Zentrum Berlin für Materialien und Energie, D-14109 Berlin, Germany}

\author{P.~Link}
\affiliation{\mbox{Forschungsneutronenquelle Heinz Maier-Leibnitz (FRM II), Technische Universität München, D-85747 Garching, Germany}}

\author{E.~Faulhaber}
\affiliation{\mbox{Forschungsneutronenquelle Heinz Maier-Leibnitz (FRM II), Technische Universität München, D-85747 Garching, Germany}}
\affiliation{Gemeinsame Forschergruppe HZB~--~TU Dresden, Helmholtz-Zentrum Berlin für Materialien und Energie, D-14109 Berlin, Germany}

\author{I.\,Glavatskyy}
\affiliation{Helmholtz-Zentrum Berlin für Materialien und Energie GmbH, Glienicker Str. 100, D-14109 Berlin, Germany}

\author{C.\,T.~Lin}
\affiliation{Max-Planck-Institut für Festkörperforschung, Heisenbergstraße 1, 70569 Stuttgart, Germany}

\author{B.\,Keimer}
\affiliation{Max-Planck-Institut für Festkörperforschung, Heisenbergstraße 1, 70569 Stuttgart, Germany}

\author{V.~Hinkov}\email[E-mail: \vspace{4pt}]{v.hinkov@fkf.mpg.de}
\affiliation{Max-Planck-Institut für Festkörperforschung, Heisenbergstraße 1, 70569 Stuttgart, Germany}

\begin{abstract}
\noindent We study the symmetry of spin excitation spectra in 122-ferropnictide superconductors by comparing the results of first-principles calculations with inelastic neutron scattering (INS) measurements on \BFCAour\ and \BFNAour\ samples that exhibit neither static magnetic phases nor structural phase transitions. In both the normal and superconducting (SC) states, the spectrum lacks the three-dimensional (3D) $4_2/m$ screw symmetry around the $(\frac{1}{\text{\protect\raisebox{0.8pt}{2}}} \frac{1}{\text{\protect\raisebox{0.8pt}{2}}} L)$ axis that is implied by the $I4/mmm$ space group. This is manifest both in the in-plane anisotropy of the normal- and SC-state spin dynamics and in the out-of-plane dispersion of the spin-resonance mode. We show that this effect originates from the higher symmetry of the magnetic Fe-sublattice with respect to the crystal itself, hence the INS signal inherits the symmetry of the unfolded Brillouin zone (BZ) of the Fe-sublattice. The in-plane anisotropy is temperature-independent and can be qualitatively reproduced in normal-state density-functional-theory calculations without invoking a symmetry-broken (``nematic'') ground state that was previously proposed as an explanation for this effect. Below the SC transition, the energy of the magnetic resonant mode \Er, as well as its intensity and the SC spin gap inherit the normal-state intensity modulation along the out-of-plane direction $L$ with a period twice larger than expected from the body-centered-tetragonal BZ symmetry. The amplitude of this modulation decreases at higher doping, providing an analogy to the splitting between even and odd resonant modes in bilayer cuprates. Combining our and previous data, we show that at odd $L$ a universal linear relationship $\hbar\Er\approx4.3\,k_{\rm B}T_{\rm c}$ holds for all the studied Fe-based superconductors, independent of their carrier type. Its validity down to the lowest doping levels is consistent with weaker electron correlations in ferropnictides as compared to the underdoped cuprates.
\end{abstract}

\pacs{74.70.Xa 78.70.Nx 75.30.Ds\vspace{-0.7em}}

\maketitle

\vspace{-5pt}\section{Introduction}\vspace{-1pt}

\vspace{-5pt}\subsection{Electronic symmetry-broken states}\vspace{-5pt}

The ground state of a paramagnetic metal naturally inherits all symmetries of its underlying crystal structure. This generally applies to the single-particle Bloch states in the periodic potential of the lattice and can be generalized to the spectra of particle-hole excitations or collective modes, such as phonons. However, several mechanisms may lead to spontaneous breaking of this crystal symmetry as the system is driven by a change of some control parameter (e.g. temperature, electron doping, or pressure) towards an ordered ground state. For such symmetry breaking to occur, both electron and lattice degrees of freedom are often required \cite{Vojta09}, as in magneto-structural or charge-density-wave \cite{Gruener88, GruenerBook, BorisenkoKordyuk08} transitions. Occasionally, though, the electron degrees of freedom alone are sufficient to lead to an instability, while the lattice only adjusts itself to the new ground state, offering little contribution to the overall energy gain \cite{Vojta09}. The most prominent examples of such electron-driven instabilities are spin-density-wave (SDW) transitions \cite{Fawcett88, Gruener94, GruenerBook}, at which a magnetic ordering wavevector is spontaneously chosen out of several equivalent Fermi surface (FS) nesting vectors, or Pomeranchuk instabilities that spontaneously lower the FS symmetry \cite{Pomeranchuk58, YamaseKohno00, YamaseKohno00a, HalbothMetzner00, OganesyanKivelson01, JakubczykMetzner09, Yamase09}.

In the special case of so-called ``\textit{electronic nematic}'' phases \cite{KivelsonFradkin98, FradkinKivelson10, SinghMazin10}, only the rotational symmetry of the electron subsystem is reduced, whereas the translational symmetry and, hence, the size of the Brillouin zone (BZ), are preserved. Such states have been extensively studied in quasi-two-dimensional compounds, such as Sr$_3$Ru$_2$O$_7$ (Ref.\,\onlinecite{BorziGrigera07}) or underdoped YBa$_2$Cu$_3$O$_{6+y}$ \cite{LeeBasov04, FujitaGoka04, HinkovHaug08, MatsudaFujita08, HaugHinkov09, HinkovKeimer10}. Recently, electronic nematic phases have been also suggested for various iron-arsenide superconductors \cite{FangYao08, XuMuller08, MazinJohannes09, BergKivelson10, KnolleEremin10, ChuangAllan10, LesterChu10, LiBroholm10, DialloPratt10, ChuAnalytis10, LeeXu10, NandiKim10} (for the latest reviews, see Refs.\,\onlinecite{Johnston10} and~ \onlinecite{LumsdenChristianson10review}).

\begin{figure}[t]
\includegraphics[width=\columnwidth]{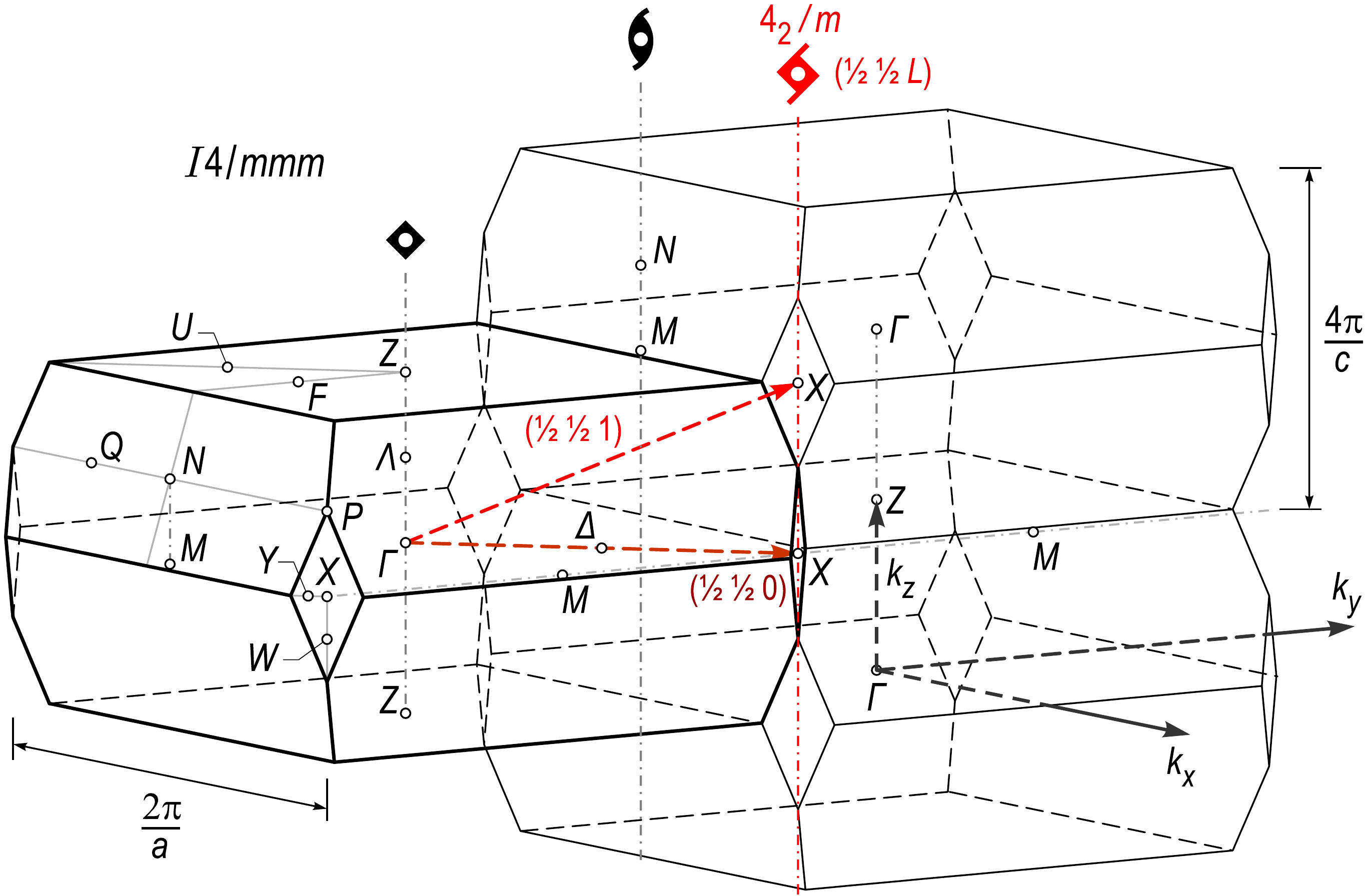}
\caption{The reciprocal-space structure of the $I4/mmm$ crystal. The BZ polyhedron of BaFe$_2$As$_2$ is drawn at the left in solid black lines, and two more such polyhedra are drawn to illustrate the 3D stacking of the Brillouin zones. Two $\protect\overrightarrow{\Gamma X}$ vectors are shown by dashed arrows: The SDW vector of the parent compound $\mathbf{Q}_\text{AFM}=(\frac{1}{\text{\protect\raisebox{0.8pt}{2}}} \frac{1}{\text{\protect\raisebox{0.8pt}{2}}} 1)$ and its in-plane projection $\mathbf{Q}_\parallel=(\frac{1}{\text{\protect\raisebox{0.8pt}{2}}} \frac{1}{\text{\protect\raisebox{0.8pt}{2}}}\kern.2pt 0)$. Symmetry axes are denoted by dash-dotted lines.}
\label{Fig:I4mmm}
\vspace{-10pt}
\end{figure}

\begin{figure}[b]
\includegraphics[width=\columnwidth]{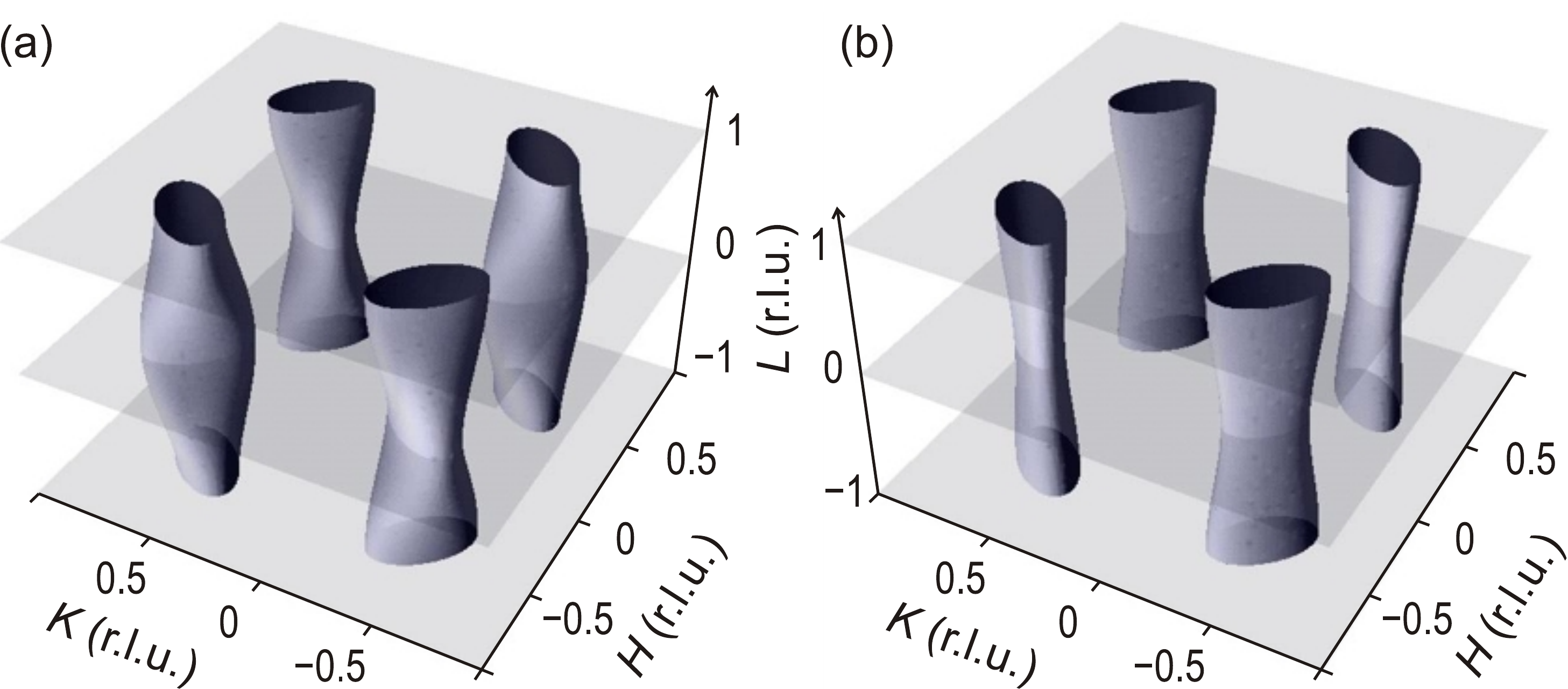}
\caption{A sketch illustrating the symmetry of spin excitations: (a) as expected from the BZ symmetry in the absence of matrix elements; (b) as actually observed experimentally in doped 122-compounds. The surfaces schematically represent constant-intensity contours of the magnetic INS response. The center of each panel corresponds to the $\Gamma$ point. Note that despite the lower symmetry in (b) due to the absence of the $4_2/m$ screw around $(\frac{1}{\text{\protect\raisebox{0.8pt}{2}}} \frac{1}{\text{\protect\raisebox{0.8pt}{2}}} L)$, the four-fold ($4/m$) rotational symmetry around $(0\,0\,L)$ is preserved.}
\label{Fig:Snakes}
\vspace{-12pt}
\end{figure}

In the following, we present the results of first-principles calculations and inelastic neutron scattering (INS) measurements of the spin excitation spectra in the normal and superconducting (SC) states of slightly under\-doped \BFNA\ (BFNA) and optimally doped \BFCA\ (BFCA) single crystals, which belong to the so-called 122-family \cite{RotterTegel08, SefatJin08, CanfieldBudko10} of ferropnictides \cite{Chu09, IshidaNakai09, RenZhao09, AswathyAnooja10, PaglioneGreene10}. Although the crystal structure of both compounds retains its body-centered-tetragonal (bct) $I4/mmm$ symmetry (Fig.\,\ref{Fig:I4mmm}) down to the lowest measurable temperatures, the low-energy spin response both in the normal and SC states has a lower symmetry in the reciprocal space \cite{DialloPratt10, LesterChu10, LiBroholm10}, as sketched in Fig.\,\ref{Fig:Snakes}, which corresponds to the unfolded BZ of the Fe-sublattice (Fig.\,\ref{Fig:BZ}). This unfolded zone is often introduced to simplify the band-structure description of the iron pnictides \cite{MazinSingh08, RaghuQi08}, but is usually considered only as a theoretical abstraction, because any realistic band structure of these systems is certainly affected by the pnictogen atoms that lower the symmetry of the direct lattice and, consequently, introduce an additional translational symmetry in the reciprocal space due to the BZ folding. Nevertheless, as we will demonstrate in the following, the absence of any appreciable magnetic moment on the pnictogen atoms allows for a much simpler description of the dynamical spin susceptibility, which experiences no structural folding and hence does not acquire the additional reciprocal-space symmetry expected in the backfolded tetragonal (structural, nonmagnetic) BZ. Therefore, as far as the magnetic fluctuations in the paramagnetic state of ferropnictides are concerned, the unfolded description of the spectrum becomes physically justified.

Due to the three-dimensional (3D) character of the 122-systems, manifest both in their electronic structure \cite{Singh08, BrouetMarsi09, VilmercatiFedorov09, LiuKondo09, MalaebYoshida09, XuHuang10, ZhangYang10, YoshidaNishi10, WangQian10} and in the substantial out-of-plane magnetic coupling in their undoped (parent) compounds \cite{EwingsPerring08, McQueeneyDiallo08, ZhaoYao08, MatanMorinaga09, ZhaoAdroja09, DialloAntropov09}, the missing symmetry operation is essentially three-dimensional, involving all three crystallographic coordinates. It corresponds to the $4_2/m$ screw symmetry around the $(\frac{1}{\text{\raisebox{0.8pt}{2}}} \frac{1}{\text{\raisebox{0.8pt}{2}}} L)$ axis \cite{FootnoteCoordinates}, shown in Fig.\,\ref{Fig:I4mmm}, and is equivalent to a product of a 90$^\circ$ in-plane rotation around the $\Gamma$ point and a translation by the reciprocal lattice vector $\mathbf{G}=\protect\overrightarrow{\Gamma\Gamma}=(1\,0\,1)$. In the following, we will show that the clear absence of such screw symmetry\,---\,a conjectured 3D analog of the electronic nematicity\,---\,can indeed be observed in the spin-excitation spectrum already in the normal (paramagnetic) state, both along the out-of-plane and along the in-plane directions of the reciprocal space. In this respect, our experimental data are in qualitative agreement with recent reports of anisotropic in-plane excitations seen both in the magnetically ordered \cite{ZhaoAdroja09, DialloAntropov09} and paramagnetic \cite{DialloPratt10, LesterChu10, LiBroholm10} states. The latter were previously associated with ``spin nematic correlations''. However, a comparison with normal-state density-functional-theory (DFT) calculations presented in section \ref{SubSec:DFT} shows good agreement between the calculated and measured susceptibilities, leading us to an alternative explanation for the lowered symmetry of the spin-excitation spectrum that does not require a symmetry-broken ground state or proximity to a quantum critical point. Instead, it turns out to be a direct consequence of the crystal structure with two Fe atoms per primitive unit cell, in which the crystalline lattice that determines the BZ geometry has a lower symmetry than its Fe-sublattice, which is responsible for the magnetism \cite{LeeVaknin10, BrownChatterji10}.

\makeatletter\renewcommand{\fnum@figure}[1]{\figurename~\thefigure.}\makeatother
\makeatletter\renewcommand{\fnum@figure}[1]{\figurename~\thefigure~(color online).}\makeatother

\vspace{-5pt}\subsection{Reciprocal space of the 122-ferropnictides}\vspace{-5pt}

\begin{figure*}[!]\newlength\LayerSpace\setlength\LayerSpace{-1.7535pt}
\includegraphics[width=0.80\textwidth]{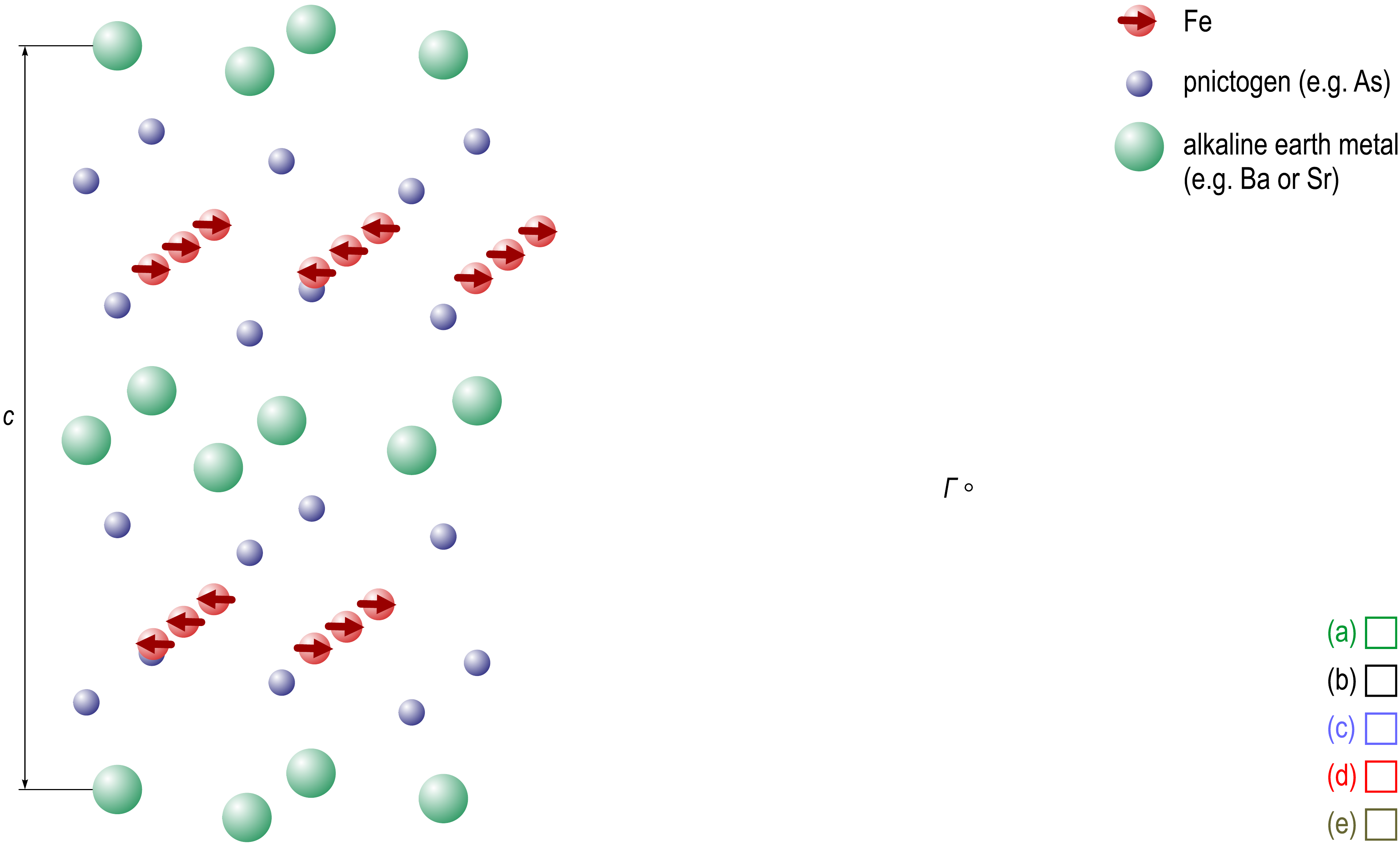}\hspace{-0.8\textwidth}\hspace{-2.422pt}
\begin{ocg}{simple tetragonal}{e}{1}\includegraphics[width=0.80\textwidth]{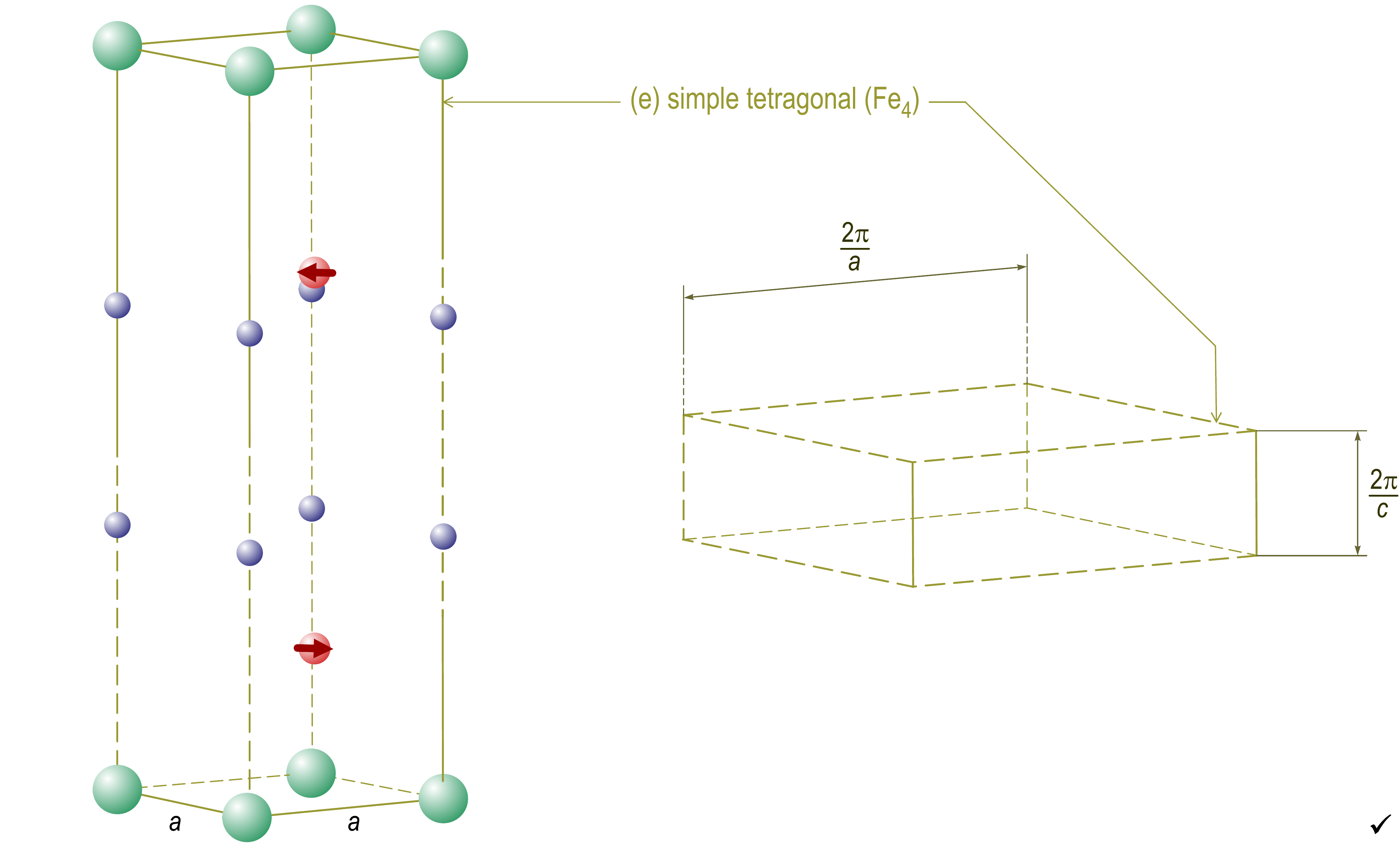}\end{ocg}\hspace{-1.2em}\raisebox{0.47em}{\ToggleLayer{simple tetragonal}{$\phantom{||||}$}}\hspace{-0.8\textwidth}\hspace{\LayerSpace}
\begin{ocg}{unfolded tetragonal}{a}{1}\includegraphics[width=0.80\textwidth]{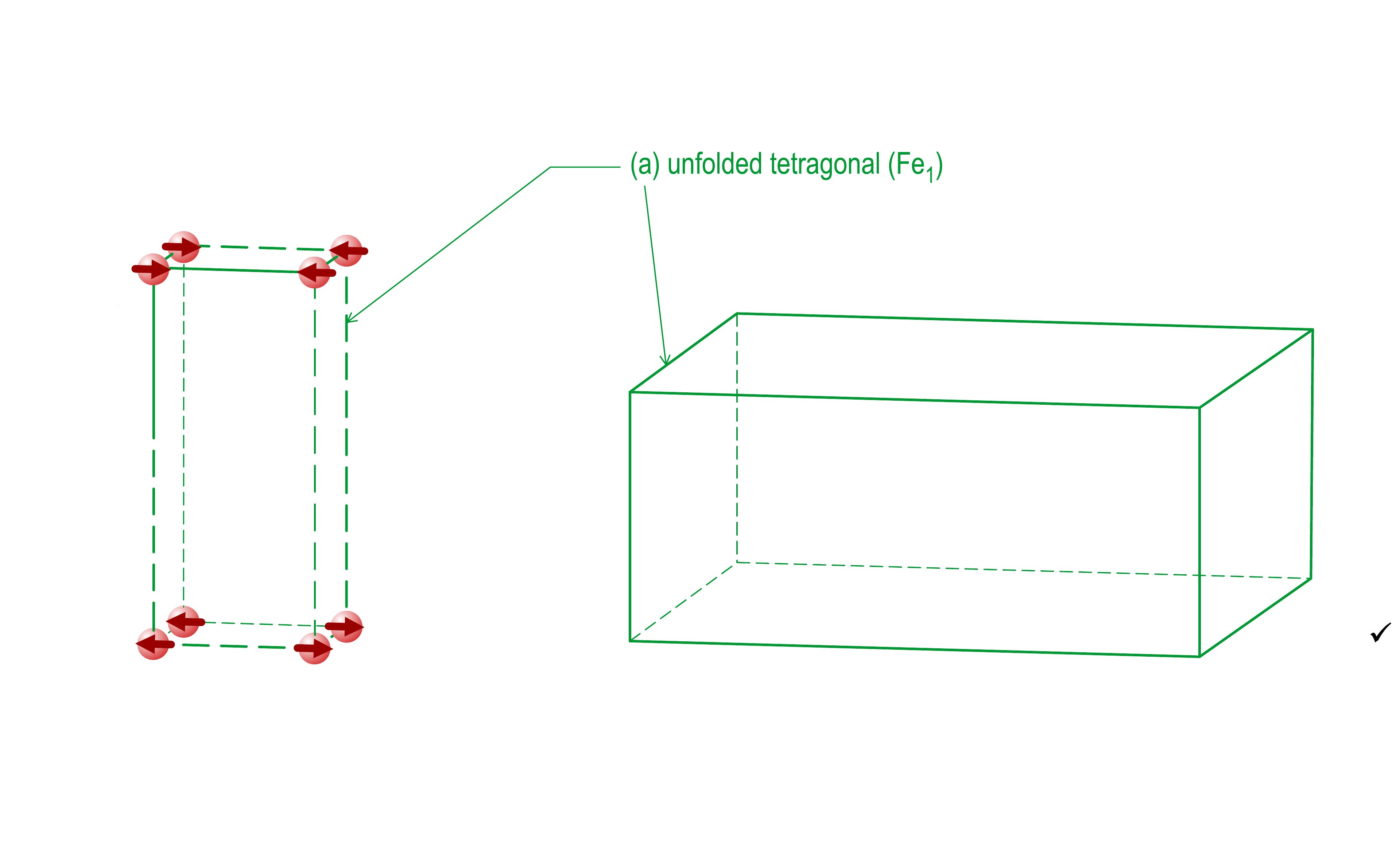}\end{ocg}\hspace{-1.2em}\raisebox{6.94em}{\ToggleLayer{unfolded tetragonal}{$\phantom{||||}$}}\hspace{-0.8\textwidth}\hspace{\LayerSpace}
\begin{ocg}{body-centered tetragonal}{b}{1}\includegraphics[width=0.80\textwidth]{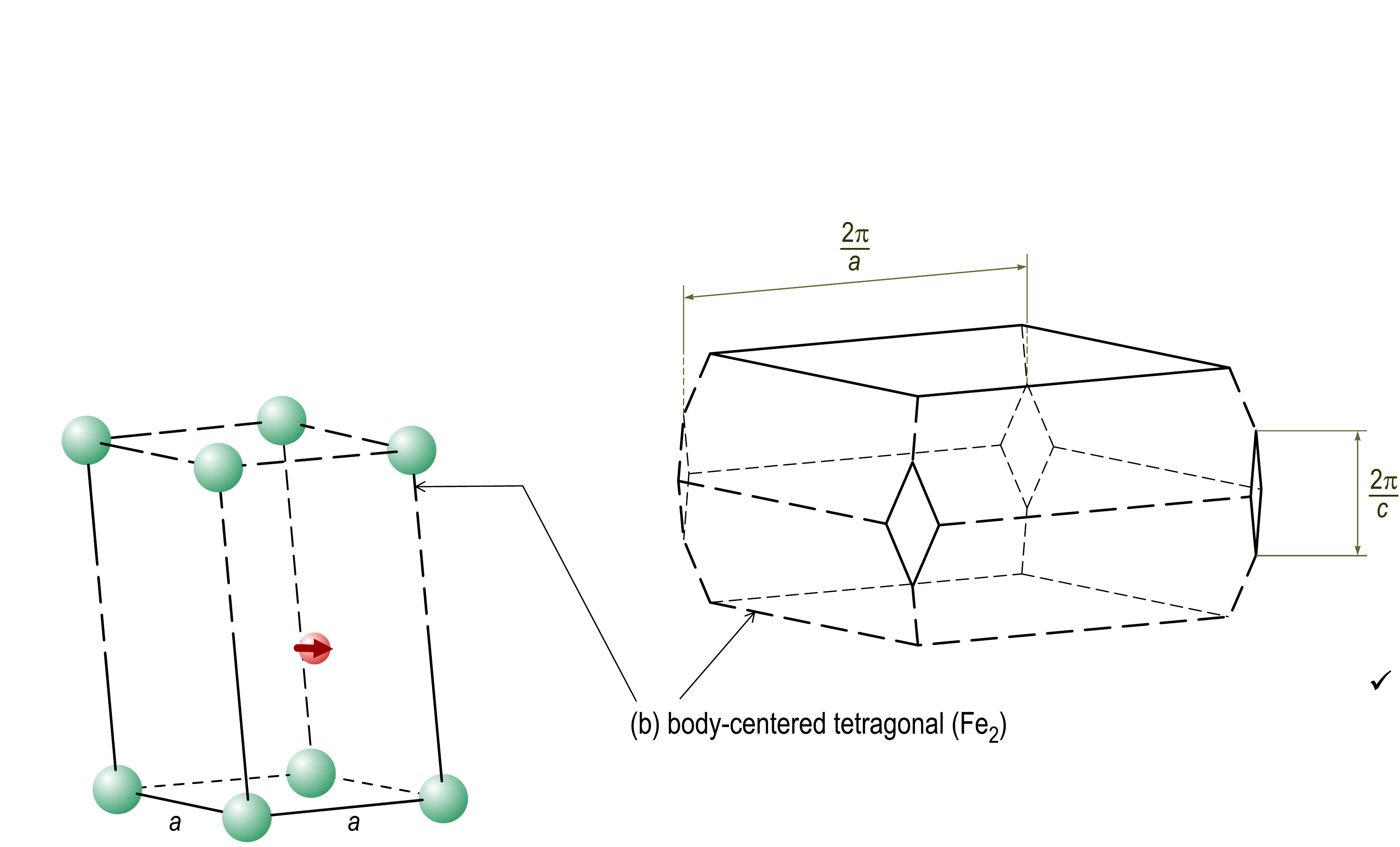}\end{ocg}\hspace{-1.2em}\raisebox{5.32em}{\ToggleLayer{body-centered tetragonal}{$\phantom{||||}$}}\hspace{-0.8\textwidth}\hspace{\LayerSpace}
\begin{ocg}{unfolded magnetic}{c}{1}\includegraphics[width=0.80\textwidth]{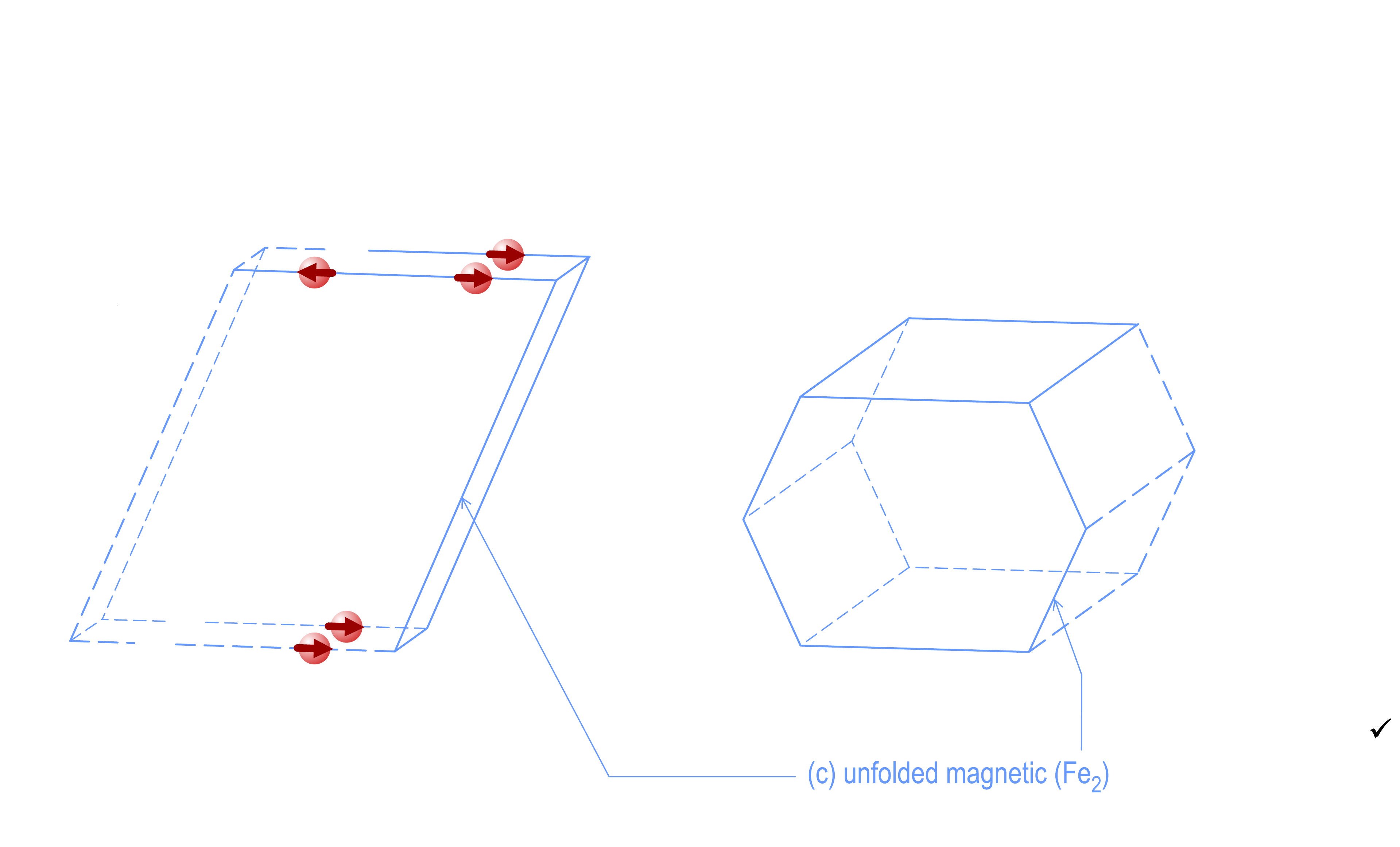}\end{ocg}\hspace{-1.2em}\raisebox{3.69em}{\ToggleLayer{unfolded magnetic}{$\phantom{||||}$}}\hspace{-0.8\textwidth}\hspace{\LayerSpace}
\begin{ocg}{doubly-folded magnetic}{d}{1}\includegraphics[width=0.80\textwidth]{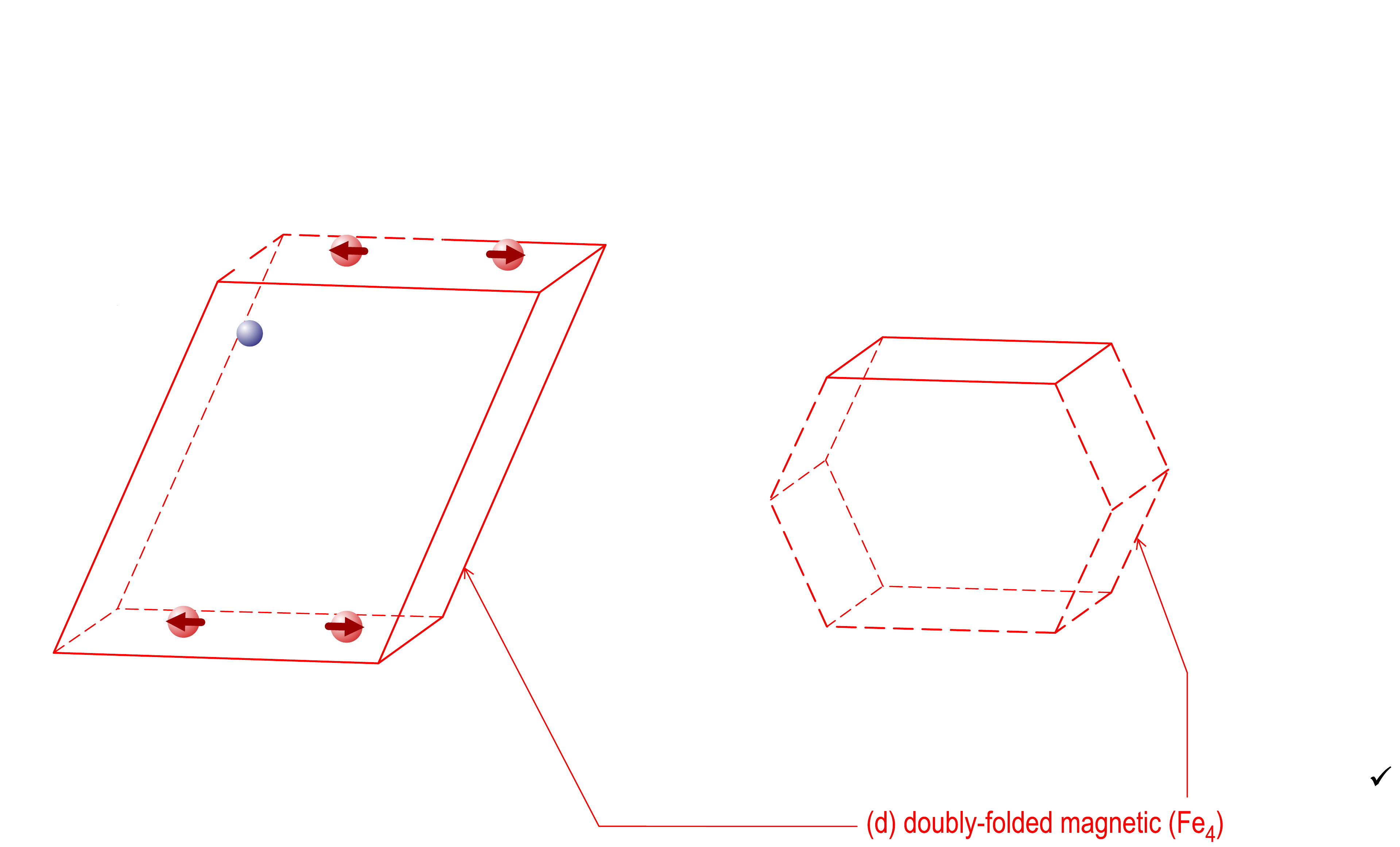}\end{ocg}\hspace{-1.2em}\raisebox{2.09em}{\ToggleLayer{doubly-folded magnetic}{$\phantom{||||}$}}\hspace{-0.8\textwidth}\hspace{\LayerSpace}
\begin{ocg}{all}{all}{1}\includegraphics[width=0.80\textwidth]{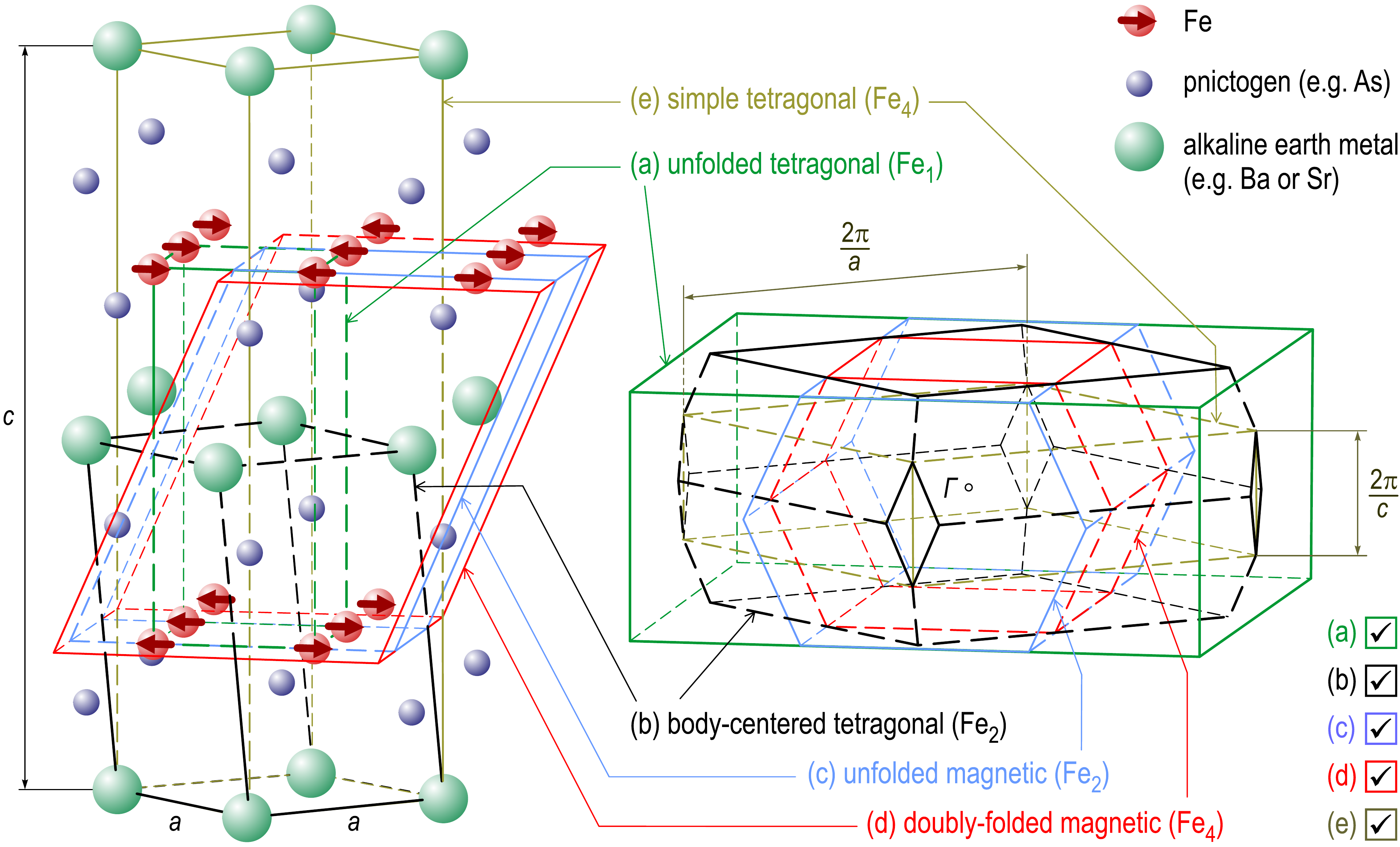}\end{ocg}
\caption{Different primitive unit cells in direct space (left) that can be introduced in 122-ferropnictides and their respective Brillouin zones (right): (a) unfolded tetragonal BZ of the Fe-sublattice with one Fe atom per unit cell (Fe$_1$); (b) structural body-centered-tetragonal BZ that corresponds to two iron atoms per primitive unit cell (Fe$_2$); (c) unfolded magnetic BZ that corresponds to the magnetically ordered Fe-sublattice in the SDW state (Fe$_2$); (d) doubly-folded magnetic BZ that results if both the lattice and magnetic structures are taken into account (Fe$_4$); (e) one of the most commonly used and experimentally convenient reciprocal-space coordinate systems that corresponds to the BZ of a simple-tetragonal direct lattice with the parameters of the real bct crystal\,---\,the notation that we also adopt for the present paper. \cite{FootnoteCoordinates} Use the checkboxes at the right to toggle the visibility of individual graphic layers.}
\label{Fig:BZ}
\vspace{-10pt}
\end{figure*}

To set the scene, in Fig.\,\ref{Fig:BZ} we summarize some of the possible coordinate systems and reciprocal-space notations that can be introduced in the 122-compounds. The figure shows five different Brillouin zones in the reciprocal space (right) and their respective primitive unit cells in direct space (left). It is natural to consider two BZ types: \textit{unfolded}, i.e. corresponding to the Fe-sublattice only, and \textit{folded}, which takes full account of the remaining nonmagnetic atoms in the unit cell. Because of the higher symmetry of the Fe-sublattice with respect to the crystal itself, the unfolded zones have twice larger volume than their folded counterparts. Next, one can also distinguish between the \textit{nonmagnetic} and \textit{magnetically folded} BZ, which correspond to the normal and SDW states, respectively. As a result, we end up with four different direct-space lattices, reciprocal-space coordinate systems, and BZ geometries that can be naturally introduced in the 122-compounds: (a) unfolded tetragonal (Fe$_1$); (b) body-centered tetragonal (Fe$_2$); (c) unfolded magnetic (Fe$_2$); (d) doubly-folded magnetic (Fe$_4$). The formulas in brackets give the number of iron atoms in the primitive unit cell. In addition, Fig.\,\ref{Fig:BZ}\,(e) shows the simple tetragonal unit cell (Fe$_4$) that defines the reciprocal-space notation used in the present paper \cite{FootnoteCoordinates}, but does not represent a primitive unit cell of the crystal. In the following, we will concentrate on the normal (paramagnetic) state, and therefore will be mainly interested in the nonmagnetic (folded or unfolded) BZ.

%

The 3D stacking of the $I4/mmm$ tetragonal Brillouin zones with the dimensions of $\frac{2\piup}{\text{\protect\raisebox{0.8pt}{$a$}}} \times \frac{2\piup}{\text{\protect\raisebox{0.8pt}{$b$}}} \times \frac{4\piup}{\text{\protect\raisebox{0.8pt}{$c$}}}$ is illustrated in Fig.\,\ref{Fig:I4mmm} and is valid both for the momentum ($\mathbf{k}$) and momentum transfer ($\mathbf{Q}$) spaces. In our notation, the quasi-two-dimensional (2D) warped hole- and electronlike FS cylinders \cite{Singh08, BrouetMarsi09, ZabolotnyyInosov09, ZabolotnyyEvtushinsky09, LiuKondo10} are centered around $\Gamma\kern-0.3pt\Lambda Z$ and $X\kern-0.3pt PX$ symmetry axes along the zone boundaries, respectively. The INS data presented in this paper were measured in the vicinity of two $\protect\overrightarrow{\Gamma X}$ wavevectors that are shown by dashed arrows: The magnetic ordering wavevector of the parent compound $\mathbf{Q}_\text{AFM}=(\frac{1}{\text{\protect\raisebox{0.8pt}{2}}} \frac{1}{\text{\protect\raisebox{0.8pt}{2}}} 1)$ and its in-plane projection $\mathbf{Q}_\parallel=(\frac{1}{\text{\protect\raisebox{0.8pt}{2}}} \frac{1}{\text{\protect\raisebox{0.8pt}{2}}}\kern.4pt 0)$. Note that the two vectors are equivalent in a tetragonal system modulo the reciprocal lattice vector $\mathbf{G}=\protect\overrightarrow{\Gamma\Gamma}=(1\,0\,1)$, because $\mathbf{Q}_\text{AFM}-\textbf{G}=(\text{--}\frac{1}{\text{\protect\raisebox{0.8pt}{2}}} \frac{1}{\text{\protect\raisebox{0.8pt}{2}}}\kern.2pt 0)\simeq(\frac{1}{\text{\protect\raisebox{0.8pt}{2}}} \frac{1}{\text{\protect\raisebox{0.8pt}{2}}}\kern.2pt 0)$. This equivalency is obliterated, however, by the magnetic order in the orthorhombic phase that selects $\mathbf{Q}_\text{AFM}$ as the preferred SDW vector. It is difficult to understand the out-of-plane component of this SDW ordering wavevector in a simple (geometric) nesting picture because of the equal nesting conditions at $\mathbf{Q}_\text{AFM}$ and $\mathbf{Q}_\parallel$, imposed by the $4_2/m$ screw symmetry. But as we will subsequently show in section \ref{SubSec:DFT}, a more rigorous calculation of the Lindhard function, taking into account the orbital matrix elements, is sufficient to resolve this dilemma.

The crystal symmetry axes are shown in Fig.\,\ref{Fig:I4mmm} by dash-dotted lines. In particular, the $4_2/m$ screw symmetry along the $X\kern-0.3pt PX$ axis appears only in the bct BZ with 2 Fe atoms per primitive cell as a result of folding, but is found neither in the unfolded BZ corresponding to the Fe-sublattice because of the missing $(1\,0\,1)$ translation, nor in the magnetic BZ because of the spontaneously broken 4-fold rotational symmetry in the SDW or orthorhombic phases (see Fig.\,\ref{Fig:BZ}). It will be especially important for our discussion because of its insensitivity to electronic twinning of the crystal, i.e. the presence of domains with different orientations of the spontaneously symmetry-broken electron states in samples with in-plane anisotropy or under the assumption of electronic nematicity. In contrast, the breaking of the 4-fold rotational symmetry around the $\Gamma\kern-0.3pt\Lambda Z$ axis cannot be directly observed, unless the sample is electronically detwinned, which can be achieved by the application of uniaxial pressure \cite{TanatarBlomberg10, ChuAnalytis10, DuszaLucarelli10} or an external magnetic field \cite{ChuAnalytis10prb}. It is also essential that the $4_2/m$ symmetry axis coincides with the $\mathbf{Q}$-space location of the spin excitations found in INS experiments, which allows us to compare the magnetic intensities along this direction. These excitations, which constitute the subject of the present study, originate from the nested hole- and electronlike Fermi surfaces \cite{MazinSingh08, RaghuQi08, Singh08, KorshunovEremin08, ChubukovEfremov08, KorshunovEremin09, BrouetMarsi09} and survive even in the overdoped regime \cite{MatanIbuka10}, i.e. well above the onset of the static SDW order in the phase diagram.

\vspace{-5pt}\subsection{Normal-state spin-excitation spectrum}\vspace{-5pt}

The normal-state spin dynamics of 122 Fe-based superconductors is dominated by an intense branch of low-energy spin fluctuations in the vicinity of the commensurate $\mathbf{Q}=(\frac{1}{\text{\raisebox{0.8pt}{2}}} \frac{1}{\text{\raisebox{0.8pt}{2}}} L)$ wavevector. It is characteristic for a nearly antiferromagnetic (AFM) metal \cite{InosovPark10} and can be well described within an itinerant framework \cite{KorshunovEremin08, MaierGraser09, EreminChubukov10, KnolleEremin10}. At higher energy transfers, the spin excitations exhibit a dispersion that has an anisotropic cross-section within every $L$\,=\,const plane. This has been evidenced in time-of-flight (TOF) experiments covering odd, even, or half-integer $L$ values \cite{LesterChu10, LiBroholm10}. The observed similarity to the magnetic parent compound \cite{ZhaoAdroja09, DialloAntropov09, DialloPratt10} served as a starting point for the proposed symmetry-broken (``electronic nematic'') ground state.

Caution has to be taken, however, since in the structural BZ (Fig.\,\ref{Fig:I4mmm}) the orthogonal $\overrightarrow{X\Gamma}$ and $\overrightarrow{XZ}$ vectors lying in the $k_xk_y$ plane (which for $L=0$ correspond to the maximal and minimal spin-wave velocities, respectively) are not equivalent. Indeed, the different shapes of the holelike barrels that alternate in a checkerboard manner, as seen in angle-resolved photoemission (ARPES) maps at a fixed excitation energy \cite{ZabolotnyyEvtushinsky09}, confirm the significance of this difference. Moreover, electronic band structure calculations within the tetragonal phase yield elliptical in-plane cross-sections of the electronlike FS sheets around the $X$ point \cite{Singh08}, which obviously do not by themselves imply any anisotropy between the $(110)$ and $(1\overline{1}0)$ directions, because the ellipse rotates by 90$^\circ$ when shifting to the next $X$ point. Therefore, discussions of the in-plane anisotropy in 122-compounds necessarily require consideration of the full 3D band structure, including the out-of-plane dispersion of the spin response along $L$. If the observed ellipticity followed the $I4/mmm$ symmetry of the crystal, then the $X$-centered intensity pattern in the spin susceptibility would be rotated by $90^\circ$ at odd $L$ with respect to even $L$ values because of the $4_2/m$ screw symmetry, as illustrated in Fig.\,\ref{Fig:Snakes}\,(a). On the contrary, the absence of this symmetry in the spin-excitation spectrum may lead to the same orientation of the ellipse at all $L$ and to the doubling of the period of intensity modulation along $(\frac{1}{\text{\raisebox{0.8pt}{2}}} \frac{1}{\text{\raisebox{0.8pt}{2}}} L)$, as shown in Fig.\,\ref{Fig:Snakes}\,(b).

In order to discriminate between these two possibilities, we performed triple-axis INS measurements in the $(H\,K\,[H\!+\!K])$ scattering plane, thus avoiding the $L$-integration that is pertinent to the TOF method. A direct comparison of the transverse and longitudinal scans around the $\smash{(\frac{1}{\text{\raisebox{0.8pt}{2}}} \frac{1}{\text{\raisebox{0.8pt}{2}}} 1})$ and $(-\frac{1}{\text{\raisebox{0.8pt}{2}}} \frac{1}{\text{\raisebox{0.8pt}{2}}}\kern.2pt 0)$ wavevectors, presented in section \ref{SubSec:Ellipticity}, shows that the excitation spectrum indeed does not fully follow the crystal symmetry, but inherits it only from the magnetically active Fe-sublattice. This consequence of the material's crystallography \textit{per~se} does not imply any spontaneously symmetry-broken states in direct space. Moreover, the vanishing $L$-dependence of the anisotropy ratio indicates that the structural contribution to the ellipticity (originating from the folded FS geometry) is not detectable within our experimental accuracy.\enlargethispage{5pt}

\vspace{-5pt}\subsection{Superconducting spin-resonance mode}\vspace{-5pt}

The magnetic resonant mode is the most prominent signature of superconductivity in the spin-excitation spectrum of several unconventional superconductors, such as single-layer \cite{HeBourges02} and bi-layer cuprates \cite{RossatMignod91, FongKeimer95, FongBourges99} or heavy-fermion systems \cite{SatoAso01}. In this respect, the Fe-based superconductors are no exception: A resonance was found in hole-doped \BKFA\ \cite{ChristiansonGoremychkin08}, optimally electron-doped BFCA and BFNA \cite{LumsdenChristianson09, ChiSchneidewind09}, underdoped BFCA \cite{ChristiansonLumsden09}, iron chalcogenides FeTe$_{1-x}$Se$_x$ \cite{QiuBao09, WenXu10, MookLumsden10a, MookLumsden10, ArgyriouHiess10, LeeXu10, LiZhang10, LumsdenChristianson10, BabkevichBendele10, LiuHu10}, and more recently in polycrystalline LaFeAsO$_{1-x}$F$_x$ \cite{WakimotoKodama10, ShamotoIshikado10}.

The resonant mode carries information about the symmetry of the SC gap\,---\,d-wave in cuprates \cite{Eschrig06} and s$_\pm$-wave in the iron arsenides \cite{MaierScalapino08, KorshunovEremin08, Mazin10}. The resonance in Fe-based superconductors shares various common aspects with cuprates, such as its abrupt intensity evolution below $T_{\rm c}$, and the fact that it is always observed at an energy \Er\ below the particle-hole continuum that sets in at twice the SC gap $\Delta$ \cite{PailhesUlrich06, SidisPailhes07}. However, there are also differences: In BFCA, the temperature evolution of \Er\ is BCS-gap-like, and no signature of a pseudogap has been found \cite{InosovPark10}.

In section \ref{SubSec:SCstate}, we will compare two further aspects of the resonant features in both systems. First, due to the intra-bilayer coupling, bilayer cuprates exhibit two resonant modes characterized by odd and even symmetries with respect to the exchange of CuO$_2$ layers within a bilayer unit, as reported for the YBa$_2$Cu$_3$O$_{6+x}$ and Bi$_2$Sr$_2$CaCu$_2$O$_{8+\delta}$ families \cite{PailhesUlrich06, SidisPailhes07, CapognaFauque07}. These modes show intensity modulations with  $L$, antiphase with respect to each other, as well as different but $L$-independent resonance energies. Although distinct resonance energies for even and odd $L$ were observed in BFNA \cite{ChiSchneidewind09, PrattKreyssig10} (see also section \ref{SubSec:Ldep}), a comparison to the cuprates has not yet been drawn, because due to the equally-spaced FeAs layers, two distinct resonant modes are not expected. Second, a linear relationship between \Er\ and $T_{\rm c}$ has been extensively discussed for cuprates, and a ratio of $\Er/k_{\rm B}T_{\rm c} \approx 5.3$ has been established for the odd resonance, for doping levels not too far from optimal \cite{SidisPailhes07}. However, progressive deviations have been noted with underdoping \cite{MookDai02, HinkovBourges07}, a violation was reported for single-layer HgBa$_2$CuO$_{4+y}$ \cite{YuLi10}, and there is an ongoing controversy about the situation in electron-doped cuprates \cite{ZhaoDai07, YuLi08}. In contrast to this, as we will show in section \ref{SubSec:DopDep}, a similar linear relationship $\Er/k_{\rm B}T_{\rm c} \approx 4.3$ is universal among all the studied Fe-based superconductors, over the entire phase diagram and independent of their carrier type, and holds down to the lowest doping levels. This means that the coupling strength (as opposed to \Er) very weakly depends on doping.

\begin{figure*}[!]\vspace{-0.6em}
\includegraphics[width=\textwidth]{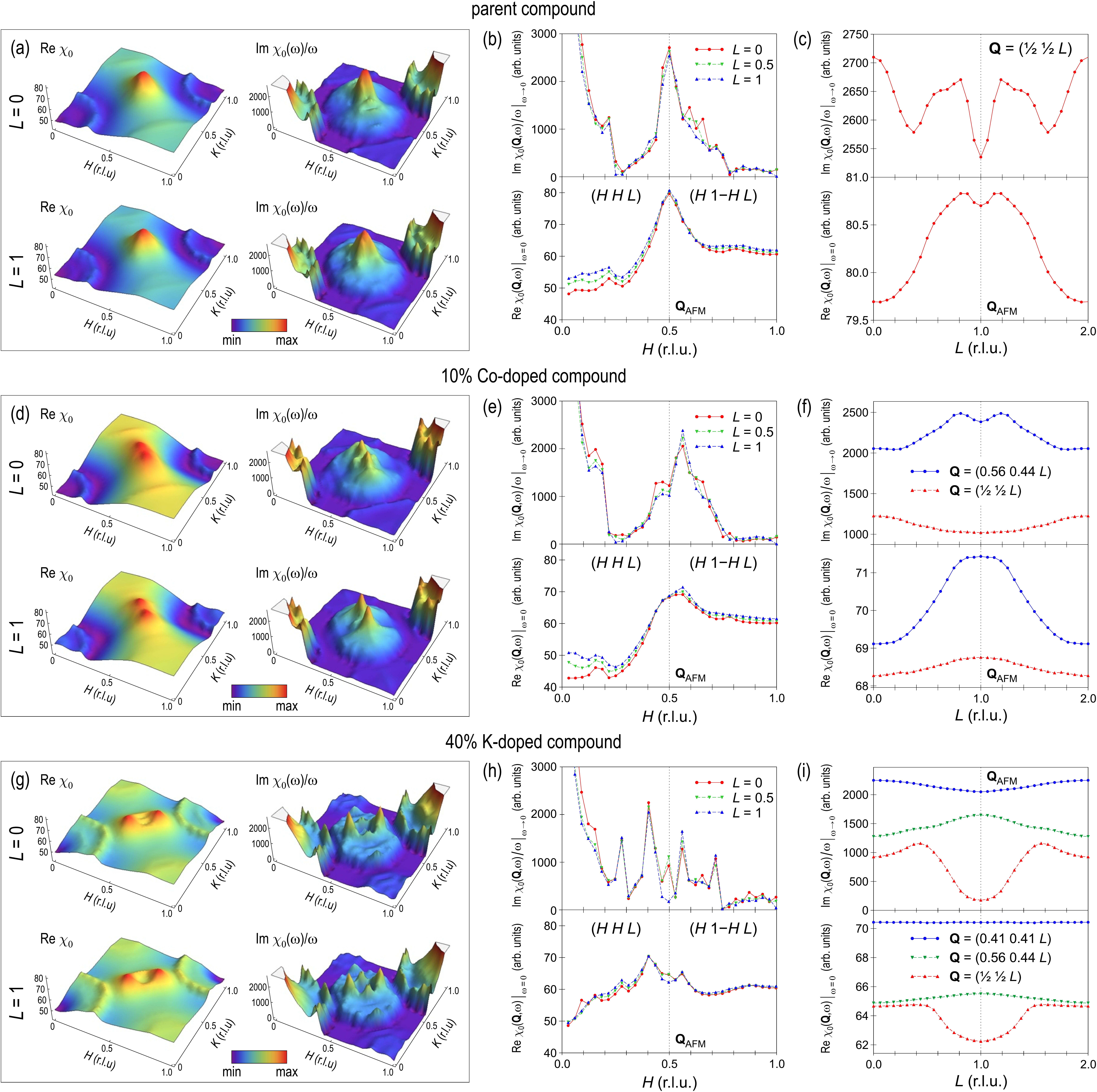}
\caption{The Lindhard function $\chi_0(\mathbf{Q},\omega)$, resulting from DFT calculations in the undoped (top), 10\% Co-doped (electron-overdoped, middle), and 40\% K-doped (optimally hole-doped, bottom) BaFe$_2$As$_2$ compounds. (a,\,d,\,g) Surface plots of the real (left) and imaginary (right) parts of the Lindhard susceptibility within the $L=0$ and $L=1$ planes. (b,\,e,\,h)~Respective profiles of $\chi_0(\mathbf{Q},\omega)$ along the high-symmetry directions, plotted at $L=0$, $1/2$, and 1. (c,\,f,\,i) $L$-dependence of $\chi_0(\mathbf{Q},\omega)$ along the $(\frac{1}{\text{\protect\raisebox{0.8pt}{2}}} \frac{1}{\text{\protect\raisebox{0.8pt}{2}}} L)$ symmetry axis and at the incommensurate peak positions (for doped compounds only).}
\label{Fig:DFT}
\vspace{-10pt}
\end{figure*}

\section{Results of first-principles calculations}

\vspace{-5pt}\subsection{Normal-state Lindhard function}\label{SubSec:DFT}\vspace{-5pt}

We start by presenting DFT calculations of the Lindhard function \cite{CallawayWang75, CallawayWang81, DresselGruener02}
\begin{eqnarray}
\chi_{0}(\mathbf{Q},\omega)=-\frac{1}{V}\sum_{\mathbf{k},n,n^{\prime}}\frac{f_{n^{\prime}}(\mathbf{k}+\mathbf{Q})-f_{n}(\mathbf{k})}{\varepsilon_{n^{\prime}}(\mathbf{k}+\mathbf{Q})-\varepsilon_{n}(\mathbf{k})+\omega+i\delta}\notag\\
\times\langle \mathbf{k},n|\hat{\sigma}_{\!+}e^{-i\mathbf{Q}\cdot\mathbf{r}}|\mathbf{k}+\mathbf{Q},n^{\prime }\rangle\langle \mathbf{k}+\mathbf{Q},n^{\prime}|\hat{\sigma}_{\!-}e^{i\mathbf{Q}\cdot \mathbf{r}}|\mathbf{k},n\rangle,\label{Eq:XiDFT}
\end{eqnarray}\vspace{2pt}
where $\varepsilon _{n}(\mathbf{k})$ is the energy of the $n$-th band, $|\mathbf{k},n\rangle$ is the corresponding wave function, $f_{n}(\mathbf{k})$ is the Fermi function, and $\hat{\sigma}_{\!\pm}$ are Pauli matrices. These calculations were performed starting from the tetragonal non-magnetic state for the experimentally determined atomic positions \cite{RotterTegel08PRB}. The chemical doping was included in the virtual crystal approximation. Further details of the calculations can be found in Ref.\,\onlinecite{YareskoLiu09}.

The surface plots of the static susceptibility $\chi_0(\mathbf{Q},\omega\rightarrow0)$ in the undoped BaFe$_2$As$_2$, 10\% Co-doped (electron-overdoped), and 40\% K-doped (optimally hole-doped) compounds are shown in Fig.\,\ref{Fig:DFT} for $L=0$ and $L=1$ together with the respective profiles along high-symmetry directions.\enlargethispage{4pt} Already in the parent compound, despite the commensurability of the nesting, a significant in-plane anisotropy of the AFM peak is observed both in the real and imaginary parts of $\chi_0$, preserving its transverse elongation at all $L$. This clearly indicates that the $4_2/m$ screw symmetry is not to be expected in the spin-fluctuation spectrum. In other words, our calculations are consistent with the lowered symmetry of the spin response that corresponds to the unfolded BZ of the Fe-sublattice, as sketched in Fig.\,\ref{Fig:Snakes}\,(b). It should be emphasized that the asymmetry of the calculated Lindhard function along the $X\Gamma$ and $XZ$ lines appears only if the matrix elements of the perturbation are properly taken into account in Eq.\,\ref{Eq:XiDFT}. If the matrix elements are neglected, $\chi_0(\mathbf{Q})$ becomes four-fold symmetric with respect to the rotation around the $(\frac{1}{\text{\raisebox{0.8pt}{2}}} \frac{1}{\text{\raisebox{0.8pt}{2}}} L)$ axis.

\begin{figure*}[t]\vspace{-0.5em}
\includegraphics[width=0.86\textwidth]{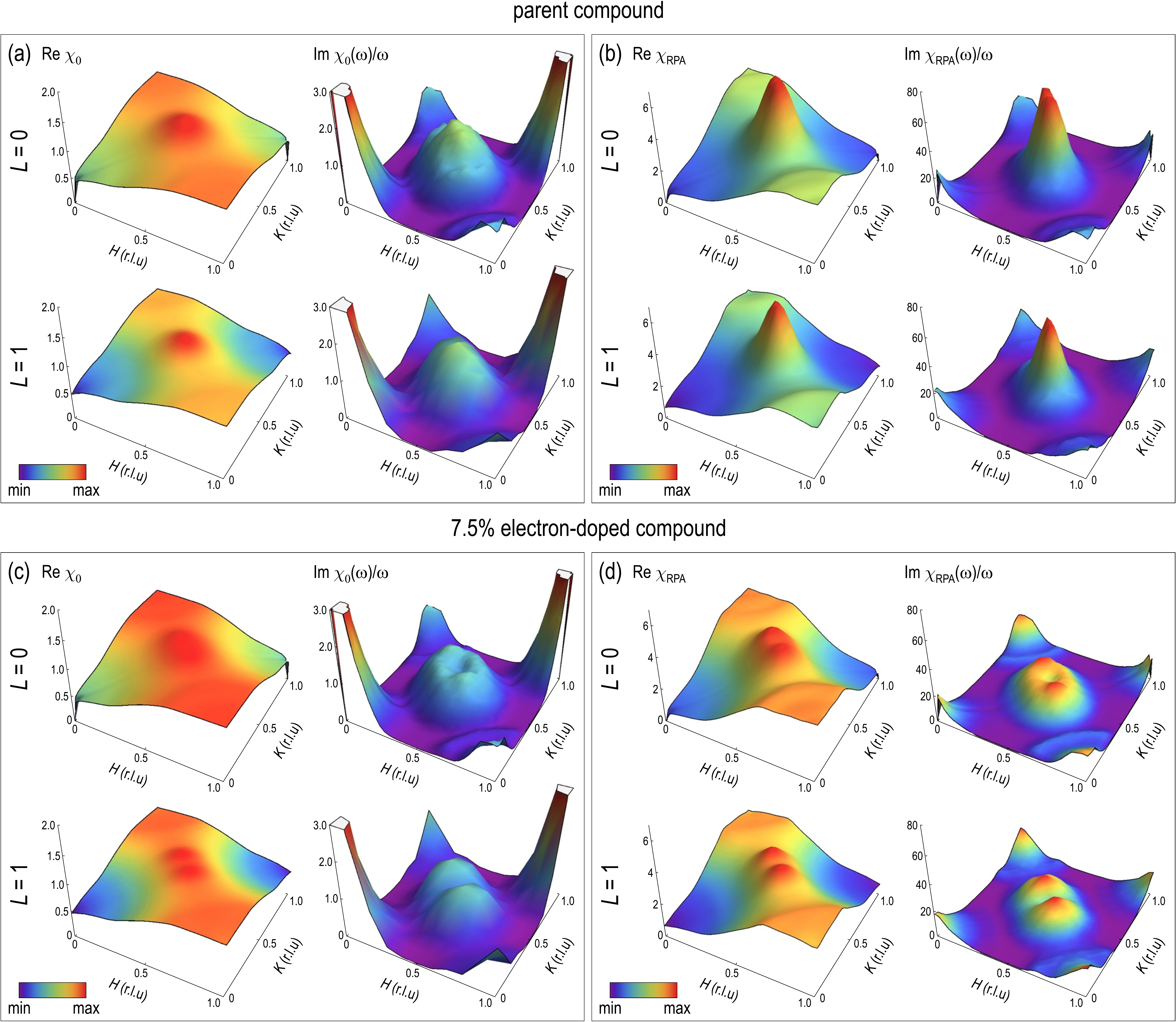}
\caption{Lindhard function (left) and renormalized RPA spin susceptibility (right) in the static limit ($\omega\rightarrow0$), calculated from a 3D tight-binding model \cite{GraserKemper10}. (a)~Lindhard function $\chi_0(H,K,0)$ and $\chi_0(H,K,1)$ for the parent (undoped) compound;
(b)~Corresponding renormalized RPA spin susceptibilities $\chi_{\rm RPA}(H,K,0)$ and $\chi_{\rm RPA}(H,K,1)$ calculated for $\overline{U}\!=\!0.8$ and $\overline{J}\!=\!0.25\,\overline{U}$. (c,\,d)~The same for 7.5\% electron-doped compound within rigid-band approximation.}
\vspace{-10pt}
\label{Fig:RPA}
\end{figure*}

The stronger response along the transverse direction is present at all $L$ values, resulting in an almost vanishing $L$-dependence, except for the weak intensity modulation that is best seen in Fig.\,\ref{Fig:DFT}\,(c). Due to this modulation, $\mathrm{Re}\chi_0(\mathbf{Q},\omega)$\,---\,the function that is responsible for the SDW instability\,---\,is $\sim$1.4\% larger at $L=1$ than at $L=0$ in undoped BaFe$_2$As$_2$, which is sufficient to explain the out-of-plane component of the 3D AFM ordering wavevector $(\frac{1}{\text{\raisebox{0.8pt}{2}}} \frac{1}{\text{\raisebox{0.8pt}{2}}} 1)$ that otherwise cannot be understood using simple geometrical nesting considerations.\enlargethispage{2pt}

As the system is doped by either electrons [Fig.\,\ref{Fig:DFT}\,(d--f)] or holes [Fig.\,\ref{Fig:DFT}\,(g--i)], the in-plane anisotropy of the $(\frac{1}{\text{\raisebox{0.8pt}{2}}} \frac{1}{\text{\raisebox{0.8pt}{2}}} L)$ peak and, consequently, the absence of the $4_2/m$ symmetry become even more apparent. The nesting peaks in the Lindhard function develop an incommensurability along the directions transverse or longitudinal to $\mathbf{Q}$, respectively, which becomes well resolved only at sufficiently high doping levels. In the Co-doped compounds below or at the optimal doping, where most of the available INS experiments were performed, the incommensurability only leads to an additional broadening of the peak in the transverse direction, and to an increase in the anisotropy ratio as compared to the undoped compound.

The $L$-dependence of $\mathrm{Re}\chi_0$ at the wave vector $(\frac{1}{\text{\raisebox{0.8pt}{2}}} \frac{1}{\text{\raisebox{0.8pt}{2}}} L)$ corresponding to stripe-like AFM correlations in the $a\kern-0.5pt b$ plane is strongly affected by doping. In undoped BaFe$_2$As$_2$, the maximum of $\mathrm{Re}\chi_0$ is found close to $L$=1, indicating that AFM correlations between Fe layers are favorable [Fig.\,\ref{Fig:DFT}\,(c)]. Electron doping suppresses the variation of the susceptibility along the $(\frac{1}{\text{\raisebox{0.8pt}{2}}} \frac{1}{\text{\raisebox{0.8pt}{2}}} L)$ line. Figure \ref{Fig:DFT}\,(f) shows, however, that the $L$-dependence at the maximum of $\mathrm{Re}\chi_0$, i.e., along the $(0.56,0.44,L)$ line, becomes more pronounced. Hole doping [Fig.\,\ref{Fig:DFT}\,(i)] leads to even stronger suppression of spin correlations with $\mathbf{Q}_{\text{AFM}}$ so that $\mathrm{Re}\chi_0$ at $(\frac{1}{\text{\raisebox{0.8pt}{2}}} \frac{1}{\text{\raisebox{0.8pt}{2}}} 1)$ becomes lower than at $(\frac{1}{\text{\raisebox{0.8pt}{2}}} \frac{1}{\text{\raisebox{0.8pt}{2}}} 0)$. The $L$ dependence at the maximum of $\mathrm{Re}\chi_0$ at $\mathbf{Q}=(0.41,0.41,L)$ is negligible, and only at the local maximum $\mathbf{Q}=(0.56,0.44,L)$, AFM correlations between the layers are still preferable.

\vspace{-5pt}\subsection{RPA spin susceptibility from a 3D tight-binding model}\label{SubSec:RPA}\vspace{-5pt}

In order to go beyond the bare spin susceptibility and account for electronic interactions, we apply the random phase approximation (RPA) to the 3D tight-binding (TB) model introduced in Ref.\,\onlinecite{GraserKemper10}, which effectively parameterizes the unfolded DFT band structure calculated for the experimental atomic positions \cite{RotterTegel08PRB}. Here the Lindhard function is calculated from the multiorbital susceptibility \cite{GraserMaier09, GraserKemper10}\vspace{-5pt}
\begin{multline}
\kern-1em(\chi_0)_{st}^{pq} ({\bf Q},\omega) = - \frac{1}{N} \sum_{{\bf k},\mu,\nu}
\frac{a_\mu^s({\bf k})\,a_\mu^p\mathstrut^*({\bf k})\,a_\nu^q({\bf k}+{\bf Q})\,a_\nu^t\kern-.5pt\mathstrut^*({\bf k}+{\bf Q})}
{ \omega + E_\nu({\bf k}+{\bf Q}) - E_\mu({\bf k}) + i 0^+} \\
\times \left[ f(E_\nu({\bf k}+{\bf Q})) - f(E_\mu({\bf k})) \right],
\end{multline}
where $p$, $q$, $s$ and $t$ are orbital indices, $\mu$ and $\nu$ label the energy dispersion $E_\nu({\bf k})$, and $f(E)$ is the Fermi function. With the summation over all momenta in the first BZ, the full 3D dispersion is taken into account. The underlying symmetry of the crystal (including the orbital composition of the bands) is reflected both in the TB band dispersions $E_\nu({\bf k})$ and in the matrix elements $a_\mu^s({\bf k})$, connecting the band and orbital spaces \cite{GraserMaier09}. Since there are indications that electronic correlations in the iron arsenide systems are moderate, as compared to the high-$T_{\rm c}$ cuprates \cite{QazilbashHamlin09, BorisKovaleva09}, we have included the Coulomb repulsion $U$ and the exchange splitting $J$ on the Fe sites in the framework of the RPA. Here the multiorbital susceptibility of the interacting system is given by \cite{GraserMaier09}
\begin{equation}
(\chi_1^{\rm RPA})_{st\mathstrut}^{pq\mathstrut} = (\chi_0)_{st\mathstrut}^{pq\mathstrut} + (\chi_1^{\rm RPA})_{uv\mathstrut}^{pq\mathstrut}\,(\widehat{U}_{\rm spin})_{wz\mathstrut}^{uv\mathstrut}\,(\chi_0)_{st\mathstrut}^{wz\mathstrut},
\end{equation}
where $\widehat{U}_{\rm spin}$ is the interaction matrix in orbital space as defined in Ref.\,\onlinecite{GraserKemper10}.
In Fig.\,\ref{Fig:RPA}, the Lindhard function\vspace{-1pt}
\begin{equation}
\chi_0(\mathbf{Q},\omega)=\frac{1}{2}\sum_{\substack{s=t\mathstrut\\p=q}}(\chi_0)_{st\mathstrut}^{pq\mathstrut}(\mathbf{Q},\omega)\vspace{-3pt}
\end{equation}
and the total RPA spin susceptibility
\begin{equation}
\chi_{\rm RPA}(\mathbf{Q},\omega)=\frac{1}{2}\sum_{\substack{s=t\mathstrut\\p=q}}(\chi_1^{\rm RPA})_{st\mathstrut}^{pq\mathstrut}(\mathbf{Q},\omega),\vspace{-3pt}
\end{equation}
calculated for $\overline{U}\!=\!0.8$ and $\overline{J}\!=\!0.25\,\overline{U}$, are shown in the static limit within $\mathbf{Q}=(HK0)$ and $\mathbf{Q}=(HK1)$ planes both for the electron-compensated parent compound and for the 7.5\% electron doping that results from a rigid-band shift of the TB bands by 33.5\,meV. The Lindhard functions presented here are not strictly equivalent to those in Fig.\,\ref{Fig:DFT}, as they are derived from independent DFT band structures and are calculated from a TB fit to the unfolded electronic bands, whereas those in Fig.\,\ref{Fig:DFT} originate directly from DFT calculations performed in the backfolded (bct) unit cell. This results in subtle differences, such as a sharper nesting peak in Fig.\,\ref{Fig:DFT}, that are not essential for the purpose of the present paper. We also note that in contrast to Ref.\,\onlinecite{GraserKemper10}, we have determined the doping level from the electron count within the tight-binding model to ensure internal consistency. The notation in Fig.~\ref{Fig:RPA} corresponds to the backfolded tetragonal BZ and therefore also differs from that of Ref.\,\onlinecite{GraserKemper10}. The RPA approach allows for a qualitative analysis of the $\mathbf{Q}$-dependence of the measured susceptibility and correctly reproduces the location of the signal in the phase space and its anisotropy. For a quantitative comparison, which is outside the scope of the present paper, approximations going beyond a standard RPA with momentum-independent interactions might be necessary.

At both doping levels, the dominant feature in $\chi_{\rm RPA}$ is located around the $\mathbf{Q}_{\rm AFM}$ wavevector, originating from the nesting of hole- and electronlike FS sheets. Its maximum appears at a nearly commensurate position in the parent compound, but the incommensurability increases drastically upon doping as a natural consequence of the rigid-band approximation. This is at variance with experiments that found a commensurate spin response in a wide range of electron doping levels \cite{InosovPark10, MatanIbuka10}. This lack of correspondence indicates that the rigid-band approximation cannot fully account for the doping effects in iron arsenides, as suggested earlier in several theoretical works \cite{Singh08, LarsonSatpathy09, Singh10, PulikkotilSchwingenschlogl10, WadatiElfimov10}.

On the other hand, the symmetry of the magnetic spectrum, as well as the tendency to larger anisotropy with increased doping, are well captured by the TB model. The Lindhard function shows good qualitative agreement with the directly calculated one from section \ref{SubSec:DFT}. The susceptibility patterns are incommensurate along the transverse direction both at $L=0$ and $L=1$, and therefore do not possess the $4_2/m$ symmetry. The RPA renormalization considerably enhances $\mathrm{Im}\,\chi_0(\mathbf{Q},\omega)/\omega$ around the nesting vector, whereas the strong peak at the $\Gamma$ point is considerably suppressed due to a much smaller Stoner factor. As a result, the overall agreement with experimental spectra that consist of a single pronounced feature centered at $(\frac{1}{\text{\raisebox{0.8pt}{2}}} \frac{1}{\text{\raisebox{0.8pt}{2}}} L)$ is further improved.

In summary, the results of our theoretical calculations indicate that the normal-state spin susceptibility contains all essential ingredients that are necessary to understand the symmetry of the measured INS spectra, both in the normal and SC states, on a qualitative level. These include both the out-of-plane modulation of the Lindhard function, peaked at the $\mathbf{Q}_{\rm AFM}$ wavevector, and the in-plane anisotropy of the nesting-driven peak, which preserves its transverse elongation at all $L$ values. Both effects lead to the absence of the $4_2/m$ screw symmetry in the spin-excitation spectrum.

\begin{figure}[t]
\includegraphics[width=\columnwidth]{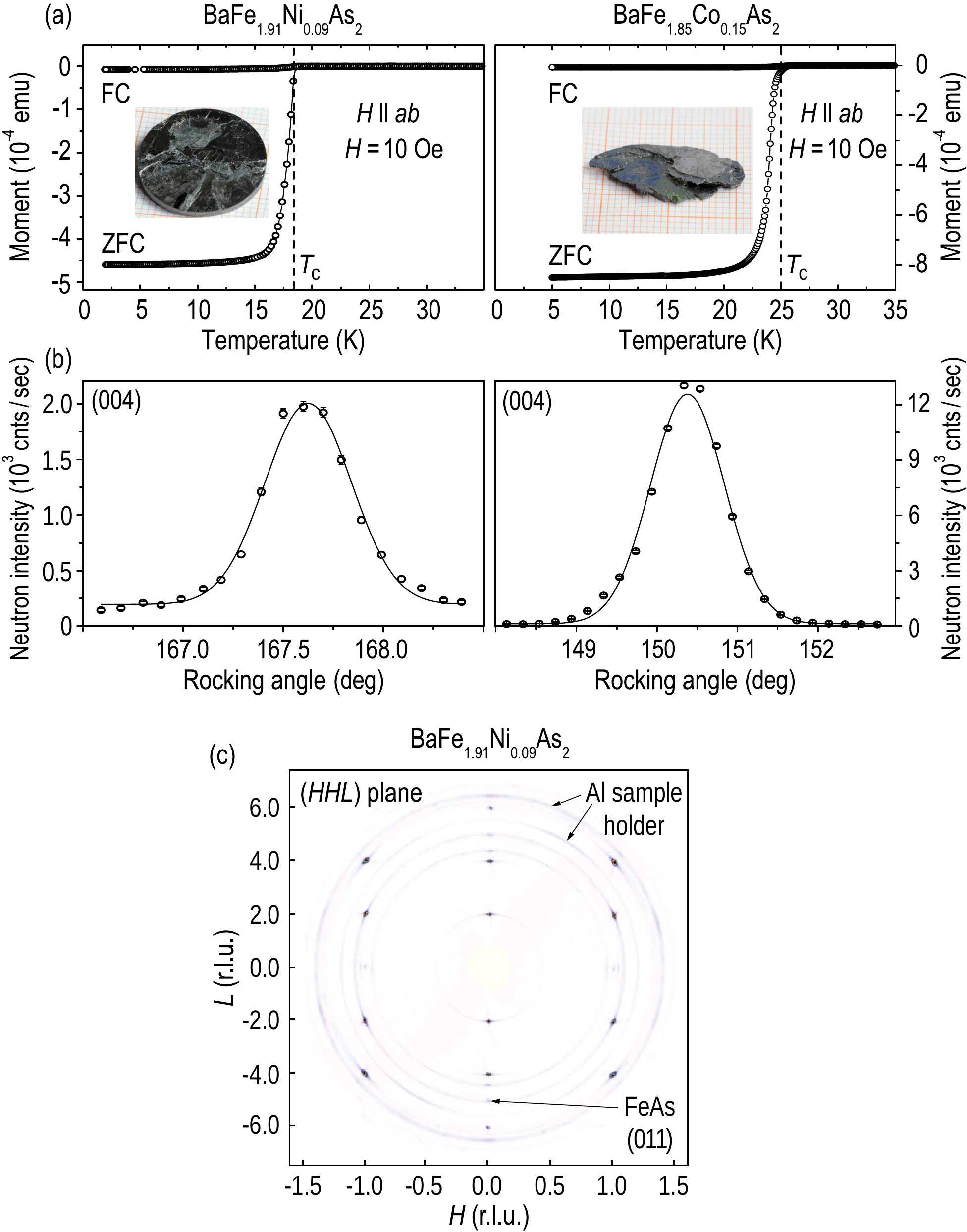}
\caption{Characterization of the samples used for the present study. (a)~Magnetization curves measured in the magnetic field of 10\,Oe, applied in-plane, after cooling in the field (FC) and in zero field (ZFC). Insets show photos of the samples. (b)~Rocking curves measured on the $(004)$ reflection in the $(HHL)$ scattering plane with a triple-axis spectrometer. (c)~Neutron diffraction pattern of the \BFNAour\ sample in the $(HHL)$ scattering plane. Powder lines coming from the Al sample holder and traces of the (Fe,Co)As flux are marked by the arrows.}
\label{Fig:Characterization}
\vspace{-12pt}
\end{figure}

\begin{figure}[b]
\includegraphics[width=\columnwidth]{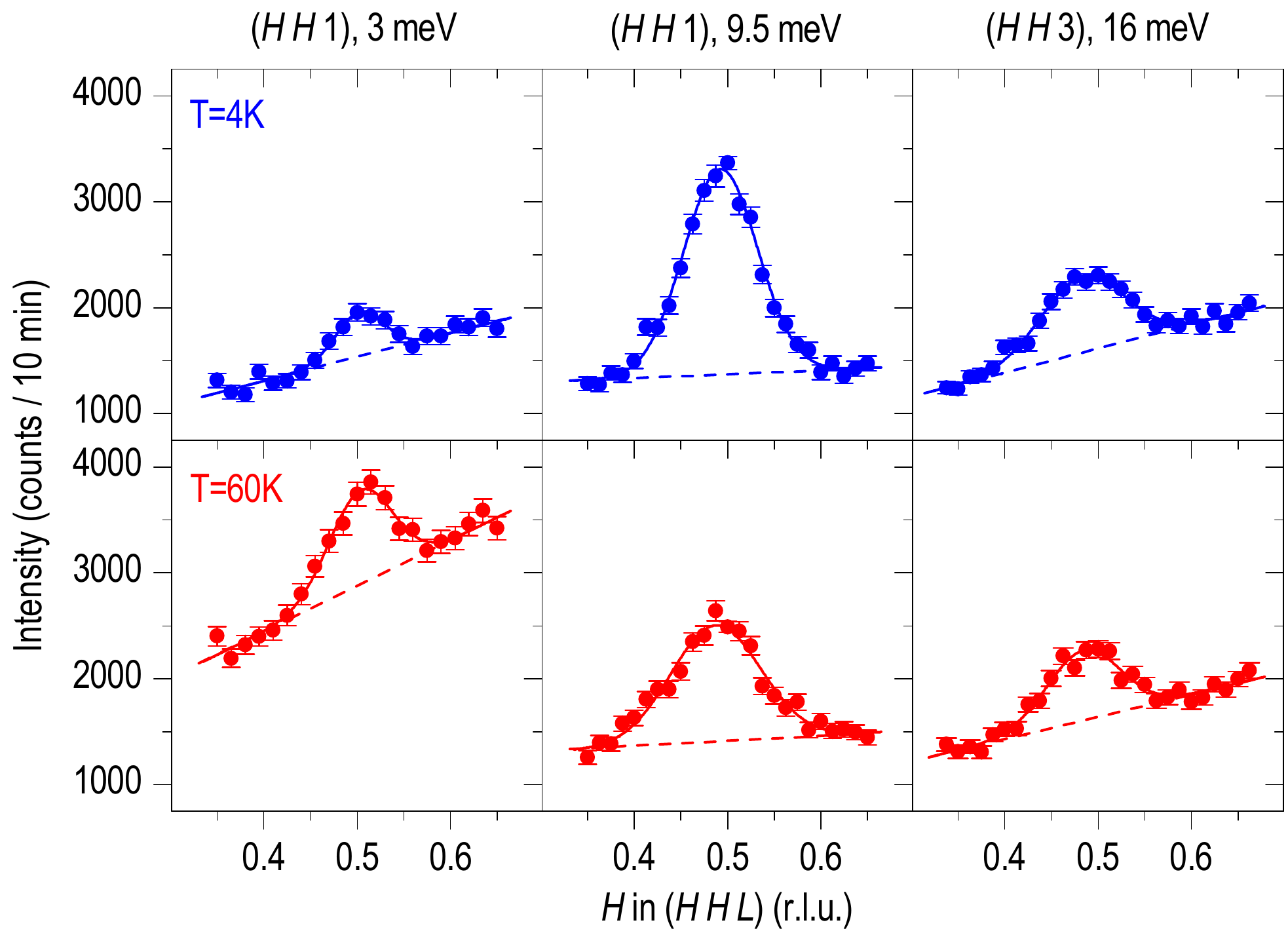}
\caption{Several raw $Q$-scans for \BFCAour, measured along the longitudinal direction in the SC state (top row, $T=4$\,K) and in the normal state (bottom row, $T=60$\,K) at three different energies: 3\,meV, 9.5\,meV, and 16\,meV. The solid lines represent Gaussian fits with a linear background. The background is indicated by dashed lines.}
\label{Fig:Q-scans}
\vspace{-12pt}
\end{figure}

\vspace{-5pt}\section{Experimental results}

\vspace{-5pt}\subsection{Sample description and experimental details}\vspace{-5pt}

\begin{figure}[t]
\includegraphics[width=\columnwidth]{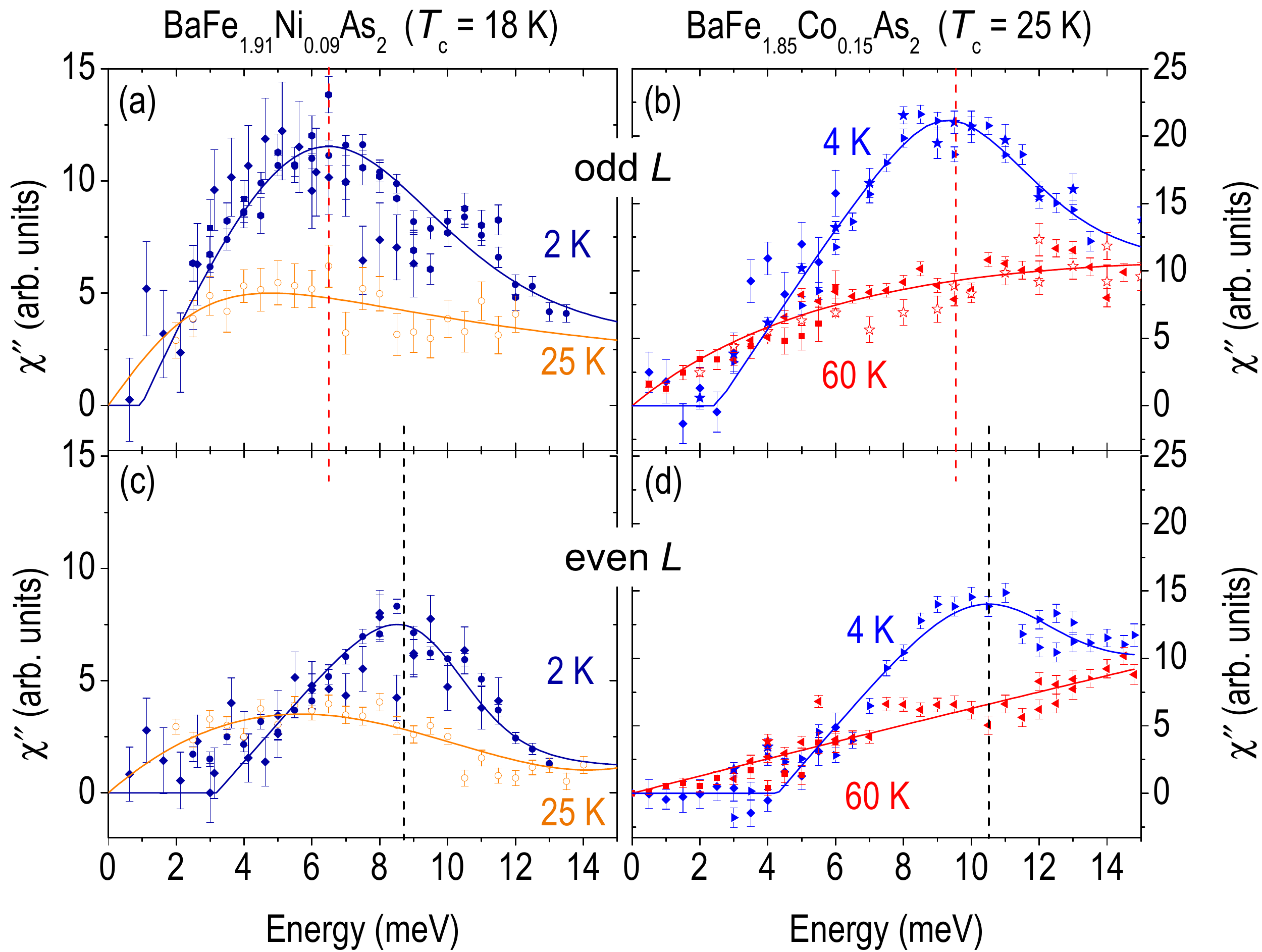}
\caption{Imaginary part of the spin susceptibility at odd (top) and even (bottom) $L$ in the normal and SC states. The left column shows data for \BFNAour\ at $\mathbf{Q}=(\frac{1}{\text{\protect\raisebox{0.8pt}{2}}} \frac{1}{\text{\protect\raisebox{0.8pt}{2}}} 1)$ and $(\frac{1}{\text{\protect\raisebox{0.8pt}{2}}} \frac{1}{\text{\protect\raisebox{0.8pt}{2}}}\kern.2pt 3)$ in (a) and at $(\smash{\frac{1}{\text{\protect\raisebox{0.8pt}{2}}} \frac{1}{\text{\protect\raisebox{0.8pt}{2}}}\kern.2pt 2})$ in (c). The right column shows corresponding data for \BFCAour. The data points were obtained from constant-$\omega$ scans and constant-$\mathbf{Q}$ scans, as described in the text. The solid lines are guides to the eye. Different symbol shapes represent data obtained in different measurements.}
\label{Fig:Normal-vs-SC}
\vspace{-12pt}
\end{figure}

\makeatletter\renewcommand{\fnum@figure}[1]{\figurename~\thefigure.}\makeatother
\begin{figure}[b]\vspace{-2pt}
\includegraphics[width=\columnwidth]{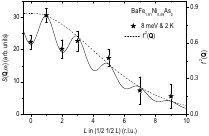}\vspace{3.2pt}
\caption{$L$-dependent magnetic intensity of \BFNAour\ in the SC state at $\mathbf{Q}=(\frac{1}{\text{\protect\raisebox{0.8pt}{2}}} \frac{1}{\text{\protect\raisebox{0.8pt}{2}}} L)$ and 8\,meV (close to the resonance energy). The dashed line shows the Fe$^{2+}\!$ spin-only magnetic form factor.}
\vspace{-10pt}
\label{Fig:Intensity}
\end{figure}

\begin{figure}[b]
\includegraphics[width=1.02\columnwidth]{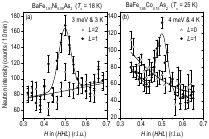}
\caption{Comparison of momentum profiles at even and odd $L$ at fixed energies that are below $\omega_\text{sg}$ for even $L$, but above it for odd $L$. (a)~\BFNAour, $T=3$\,K and $\omega=3$\,meV. (b)~\BFCAour, $T=4$\,K and $\omega=4$\,meV.}
\vspace{-10pt}
\label{Fig:GapDispersion}
\end{figure}
\makeatletter\renewcommand{\fnum@figure}[1]{\figurename~\thefigure~(color online).}\makeatother

The single crystals of \BFNAour\ ($T_{\rm c}=18$\,K, $m\approx4$\,g) and \BFCAour\ ($T_{\rm c} = 25$\,K, $m\approx1$\,g) were grown by the FeAs-flux method \cite{SunLiu09, LiuSun10} and characterized by energy-dispersive x-ray analysis, SQUID magnetometry, and single-crystal neutron diffraction using the E2 flat-cone diffractometer at the Helmholtz-Zentrum Berlin für Materialien und Energie. Magnetization measurements on several small pieces of each sample revealed sharp SC transitions at $T_{\rm c}=18$\,K and 25\,K, respectively, as shown in Fig.\,\ref{Fig:Characterization}\,(a). Both in the $(HHL)$ and $(HK0)$ planes, neutron diffraction patterns exhibit well defined Bragg spots with narrow mosaicity $< 1^\circ$ [Fig.\,\ref{Fig:Characterization}\,(b)] and no signatures of multiple single-crystalline grains, but with some polycrystalline contamination originating both from the main phase and to a lesser extent from traces of the (Fe,Co)As flux [see Fig.\,\ref{Fig:Characterization}\,(c)]. We therefore had to optimize the scattering conditions in our INS measurements by avoiding the appearance of spurious inelastic peaks caused by such contamination. No structural or SDW transitions were detected down to 2\,K in both samples, consistent with the known phase diagrams \cite{ChuAnalytis09, LesterChu09, CanfieldBudko09, NandiKim10, LiuSun10, NiThaler10}.

The INS measurements were performed at the triple-axis spectrometers PANDA and PUMA (FRM-II, Garching), IN8 (ILL, Grenoble), and 2T (LLB, Saclay). The instruments were operated in their high-flux setup without collimators, using focussed pyrolytic-graphite (002) monochromators and analyzers. Measurements were done in the constant-$k_{\rm f}$ mode, with $k_{\rm f}$ = 1.55\,{\AA}$^{-1}$ ($E_{\rm f}=4.98$\,meV) or $k_{\rm f}$ = 2.662\,{\AA}$^{-1}$ ($E_{\rm f}=14.7$\,meV). Correspondingly, either a cold Be-filter or two pyrolytic-graphite filters were used for higher-order neutron elimination.

The data for the present work were collected in the $(HHL)$ and $(H\,K\,[H\!+\!K])$ scattering planes. Throughout this paper we are using backfolded tetragonal notation \cite{FootnoteCoordinates}, in which $\mathbf{Q}_\text{AFM}=(\frac{1}{\text{\raisebox{0.8pt}{2}}} \frac{1}{\text{\raisebox{0.8pt}{2}}} 1)$ corresponds to the AFM ordering wavevector of the parent compound. We quote the wavevector $\mathbf{Q}=(HKL)$ in reciprocal lattice units (r.\,l.u.), i.\,e. in units of the conventional reciprocal lattice vectors $\textbf{a}^*$, $\textbf{b}^*$, and $\textbf{c}^*$ ($a^*$ = $b^*$ = $2\piup/a$, $c^*$ = $2\piup/c$) that would correspond to a simple tetragonal unit cell with the same dimensions. The room-temperature lattice constants are $a = b = 3.94$\,{\AA}, $c = 12.86 $\,{\AA} for \BFNAour\ and $a = b = 3.92$\,{\AA}, $c = 12.84$\,{\AA} for \BFCAour. For the sake of a compact notation we will set $\hbar=1$ in the following and quote the energy transfer $\omega$ in meV.

\vspace{-5pt}\subsection{$L$-dependence of the INS spectra}\label{SubSec:Ldep}\vspace{-5pt}

In Fig.\,\ref{Fig:Q-scans}, several representative longitudinal $Q$-scans across the AFM wavevector are shown \cite{InosovPark10}. One can see that both in the normal and SC states, the signal is well fitted by a single Gaussian peak with a linear background, showing no signatures of incommensurability along this reciprocal-space direction within the low-energy range of up to $\sim2\Delta$. In Fig.\,\ref{Fig:Normal-vs-SC}, we show the energy dependence of the experimentally measured imaginary part of the spin susceptibility \x\ at $\mathbf{Q}=(\frac{1}{\text{\raisebox{0.8pt}{2}}} \frac{1}{\text{\raisebox{0.8pt}{2}}} L)$ for both samples at even and odd $L$, obtained from the raw INS data after background subtraction and Bose-factor correction. The measured signal has also been corrected to account for the energy-dependent fraction of higher-order neutrons. The data were acquired by performing a series of full  $\mathbf{Q}$-scans similar to those shown in Fig.\,\ref{Fig:Q-scans} at different fixed energies and an energy scan at $\mathbf{Q}=(\frac{1}{\text{\raisebox{0.8pt}{2}}} \frac{1}{\text{\raisebox{0.8pt}{2}}} L)$. To estimate the background for the latter, we used a linear interpolation for the background obtained from Gaussian fits to the full $\mathbf{Q}$-scans, or measured points appropriately offset to both sides from $(\frac{1}{\text{\raisebox{0.8pt}{2}}} \frac{1}{\text{\raisebox{0.8pt}{2}}} L)$. The error bars correspond to one standard deviation of the neutron count and do not include the normalization errors. The two left panels of Fig.\,\ref{Fig:Normal-vs-SC} show data on \BFNAour, measured in the SC and normal states at $L=1$ and 3 [panel (a)] and at $L=2$ [panel (c)]. The respective data for \BFCAour\ are shown at the right.

Already in the normal state, a difference between odd and even $L$ values can be observed. For both samples, the normal-state spectral weight, integrated over $\textbf{Q}$ and $\omega$ up to 14\,meV, is $\sim$\,60\% larger at odd than at even $L$. Such a difference cannot be a consequence of the magnetic form factor, which would be smaller at $L=1$ than at $L=0$, producing the opposite effect. On the other hand, this difference is reminiscent of the SDW phase of the parent compounds, where low-energy magnon branches are present only near magnetic Bragg peaks at odd $L$, whereas spin waves at even $L$ are gapped and thus yield zero intensity at low energy \cite{McQueeneyDiallo08, ZhaoYao08, MatanMorinaga09, ZhaoAdroja09, DialloAntropov09}. However, in the paramagnetic state, the normal-state intensity at even $L$ is only moderately suppressed [cf. Fig.\,\ref{Fig:DFT}\,(c) and (f)]. Here we note that the absence of any magnetic Bragg intensity at $(\frac{1}{\text{\raisebox{0.8pt}{2}}}\,-\!\frac{1}{\text{\raisebox{0.8pt}{2}}}\,0)$ or $(-\frac{1}{\text{\raisebox{0.8pt}{2}}}\, \frac{1}{\text{\raisebox{0.8pt}{2}}}\,0)$ in the SDW state is fully consistent with the unfolded-BZ scheme. Indeed, as can be seen from Fig.\,\ref{Fig:BZ}, these two $X$ points correspond to the zone center in the doubly-folded magnetic BZ, which means that the influence of the As superstructure would lead to an appearance of magnetic Bragg-peak replicas at these points. In a twinned crystal, this would imply equivalency of all the $(\pm\frac{1}{\text{\raisebox{0.8pt}{2}}} \pm\frac{1}{\text{\raisebox{0.8pt}{2}}}\, L)$ points up to the magnetic structure factor. The fact that these replicas have not been observed by neutron diffraction indicates that the structure factor for the As-superstructure reflections is negligibly small or zero. In other words, no folding of the magnetic signal occurs due to the As sublattice, and hence the unfolded-BZ scheme is perfectly justified. Our results presented in this and the following sections serve to generalize these arguments to the inelastic magnetic signal.

At first, we consider the low-temperature spectra that exhibit the SC resonant mode. We define the resonance energy \Er\ as the maximum of \x\ in the SC state and discriminate between its value at even and odd $L$, \Ereven\ and \Erodd, where necessary. The dashed vertical lines mark these positions for odd and even $L$ in the upper and lower panels of Fig.\,\ref{Fig:Normal-vs-SC}, respectively. We note that the resonance energies at odd and even $L$ differ by more than 2\,meV in \BFNAour, in agreement with Ref.\,\onlinecite{ChiSchneidewind09}. In contrast, this difference is only about 1\,meV in optimally doped \BFCAour, as seen in Fig.\,\ref{Fig:Normal-vs-SC}~(c) and (d).

Fig.\,\ref{Fig:Intensity} shows the magnetic intensity evolution near the resonance energy along $L$, obtained from Gaussian fits of full constant-energy scans around $\mathbf{Q}=(\frac{1}{\text{\raisebox{0.8pt}{2}}} \frac{1}{\text{\raisebox{0.8pt}{2}}} L)$ at 8 meV. Similarly to the normal-state intensity, it is modulated periodically in $L$ (up to the magnetic form factor), analogous to the magnons in the parent compound. Two factors can be responsible for the observed modulation. First, as the normal-state intensity is already lower at even $L$, it will preserve this modulation after redistribution of the spectral weight due to the opening of the SC gap below $T_\text{c}$. Second, the higher energy of the resonance at even $L$ is closer to (or even within) the particle-hole continuum, which may result in stronger damping and additional intensity reduction.

Not only \Er\ and the spectral weight of the resonance, but also the energy range below the resonance peak that is depleted upon entering the SC state (which we refer to as the \textit{SC spin gap}) depends on $L$. We define the spin-gap energy $\omega_\text{sg}$ as the intersection of the low-energy linear extrapolation of \x\ at 2 or 4\,K with the $\chi''=0$ line (Fig.\,\ref{Fig:Normal-vs-SC}). Inspection of Fig.\,\ref{Fig:GapDispersion}, where we compare constant-energy scans at even and odd $L$ for both samples, clearly shows that $\omega_\text{sg}$ is larger at even $L$. We remark that the SC spin gap should not be mistaken for the SC gap $\Delta$ to which it is only indirectly related: $\omega_\text{sg}$ is determined by the energy, \Er, and the width of the resonant mode.

Recalling that the $X$ points in the BZ for odd and even $L$ values are equivalent due to the above-mentioned screw symmetry, we can now conclude that this symmetry is absent in the spin-excitation spectra of both samples based on the following evidence observed in the out-of-plane direction: (i)~different normal-state intensities at odd and even $L$; (ii)~different resonance energies \Erodd\ and \Ereven; (iii)~periodic $L$-dependent intensity of the resonance; (iv)~the corresponding difference of the spin gaps $\omega_{\rm sg,odd}$ vs. $\omega_{\rm sg,even}$.

\begin{figure}[!]
\includegraphics[width=0.982\columnwidth]{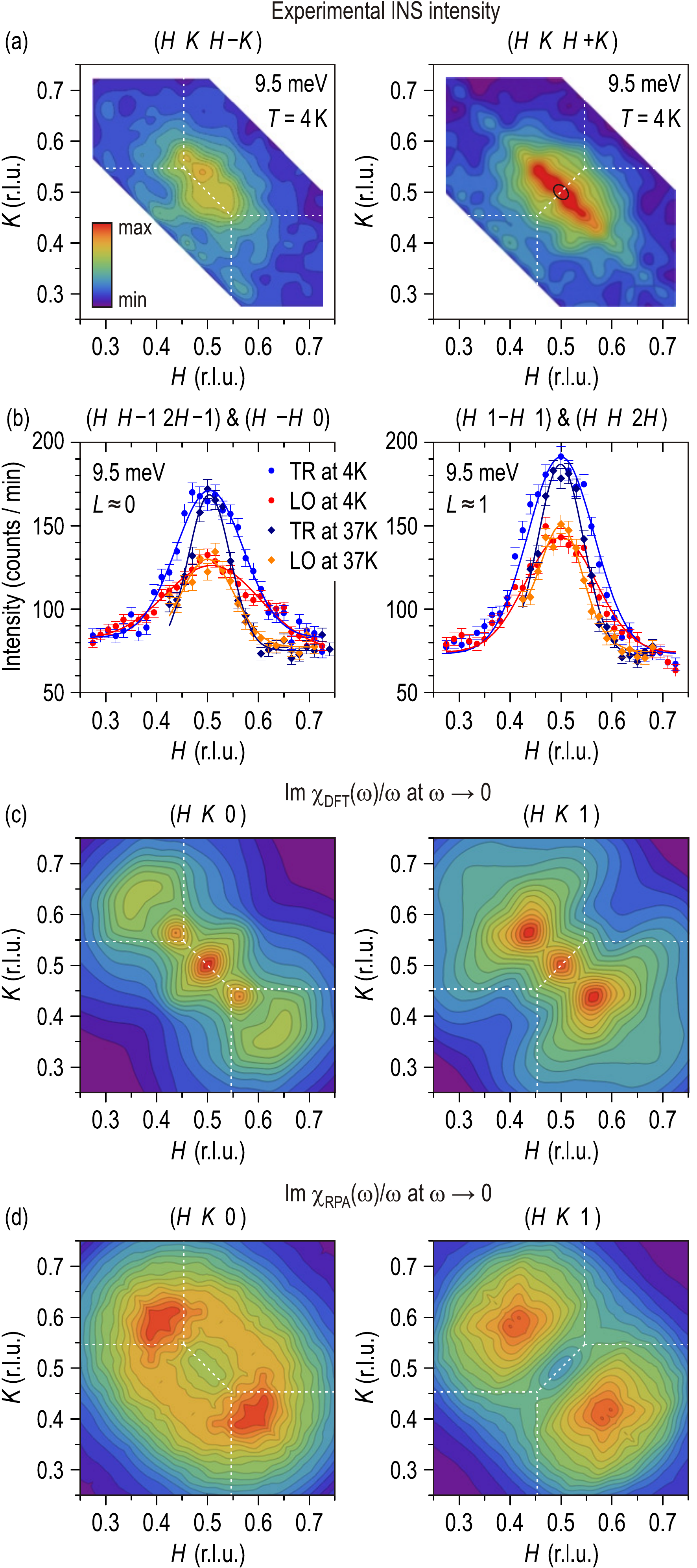}
\caption{(a)~Experimental intensity distributions for \BFCAour\ near $\mathbf{Q}_\parallel$ (left) and $\mathbf{Q}_{\rm AFM}$ (right), measured in the $(H\,K\,[H+K])$ scattering plane in the SC state ($T=4$\,K) at the resonance energy (9.5\,meV). The small black ellipse around $(\frac{1}{\text{\protect\raisebox{0.8pt}{2}}} \frac{1}{\text{\protect\raisebox{0.8pt}{2}}} 1)$ is a 9.5\,meV cross-section of the spin wave dispersion for the CaFe$_2$As$_2$ parent compound \cite{ZhaoAdroja09, DialloAntropov09}, shown for comparison. The white dotted lines are BZ boundaries. (b)~Comparison of the longitudinal (LO) and transverse (TR) cross-sections of the data from panel (a) around $L=0$ (left) and $L=1$ (right). (c)~The Lindhard function ${\rm Im}\chi_{\rm DFT}(\omega)/\omega$ at 7.5\% Co-doping, calculated by DFT in the same reciprocal space regions. (d)~The same for the RPA-renormalized low-energy spin susceptibility ${\rm Im}\chi_{\rm RPA}(\omega)/\omega$ [same as in Fig.\,\ref{Fig:RPA}\,(d)], calculated from a TB model in the rigid-band approximation.}
\label{Fig:Ellipticity}
\end{figure}

\vspace{-5pt}\subsection{In-plane anisotropy of the spin susceptibility}\label{SubSec:Ellipticity}\vspace{-5pt}

Another piece of evidence for the lowered reciprocal-space symmetry is associated with the in-plane anisotropy of the measured INS intensity \cite{ZhaoAdroja09, DialloAntropov09, DialloPratt10, LesterChu10, LiBroholm10}. In Fig.\,\ref{Fig:Ellipticity}\,(a), we show experimental constant-energy maps, interpolated from a series of triple-axis $\mathbf{Q}$-scans in the vicinity of $\smash{(\frac{1}{\text{\raisebox{0.8pt}{2}}} \frac{1}{\text{\raisebox{0.8pt}{2}}} 1})$ and $(-\frac{1}{\text{\raisebox{0.8pt}{2}}} \frac{1}{\text{\raisebox{0.8pt}{2}}}\kern.2pt 0)$ wavevectors, measured in the $(H\,K\,[H\!+\!K])$ scattering plane. We compare them with the calculated dynamic spin susceptibilities of the paramagnetic tetragonal phase, plotted in the equivalent regions of $\mathbf{Q}$-space surrounding the $X$ points. Panel (c) shows the imaginary part of the Lindhard function $\mathrm{Im}\,\chi_0(H,K,0)$ (left) and $\mathrm{Im}\,\chi_0(H,K,1)$ (right) in the vicinity of $\mathbf{Q}_\parallel$ and $\mathbf{Q}_{\rm AFM}$, respectively, for 7.5\% Co-substitution, as calculated by DFT in the virtual crystal approximation. Panel (d) displays the respective results for the RPA-enhanced susceptibility $\mathrm{Im}\,\chi_{\rm RPA}(\piup+q_x,\piup+q_y,0)$ and $\mathrm{Im}\,\chi_{\rm RPA}(\piup+q_x,\piup+q_y,1)$ [same as in Fig.\,\ref{Fig:RPA}\,(d)], calculated in the rigid-band approximation from the TB model \cite{GraserKemper10} at 7.5\% electron doping.

Notably, the transverse elongation of the susceptibility pattern is preserved at all $L$ values both in the measured INS signal and in the results of both calculations, meaning that the longer axis of the ellipse is oriented either along $\overrightarrow{XZ}$ or along $\overrightarrow{X\Gamma}$ directions for even and odd $L$, respectively. This anisotropy is insensitive to the SC transition and persists also in the normal state. Neither the widths of the peaks nor their anisotropy experience any change across $T_{\rm c}$ within our experimental accuracy, as evidenced by Fig.\,\ref{Fig:Ellipticity}\,(b).

In comparison to the magnetically ordered parent compound \cite{FootnoteParent}, which exhibits a steep spin wave dispersion cone around $\mathbf{Q}_{\rm AFM}$, as shown by a small black ellipse in Fig.\,\ref{Fig:Ellipticity}\,(a), electron doping tends to increase the transverse incommensurability [cf.~Fig.\,\ref{Fig:DFT}\,(c,\,d)] and, in addition, leads to softening of spin excitations predominantly in the transverse direction \cite{LesterChu10}. This results in a rapid increase of the anisotropy ratio with increasing doping. The emerging pattern resembles the ``unusual quasi-propagating excitations'' observed at higher energies in a similar compound by Li \textit{et al.} \cite{LiBroholm10}, as well as the pair of incommensurate peaks seen in FeTe$_{1-x}$Se$_x$ (Refs.\,\,\onlinecite{MookLumsden10, ArgyriouHiess10, LeeXu10, LumsdenChristianson10, BabkevichBendele10}). In the light of our present results, the former can be understood as two incommensurate branches of itinerant Stoner-like excitations, driven by FS nesting, as in the case of iron chalcogenides \cite{ArgyriouHiess10, LeeXu10, LiZhang10, LiuHu10}. The fact that such incommensurability has not been resolved experimentally at low energies is not surprising, because for sufficiently small doping levels at which the overwhelming majority of INS experiments was performed, the two incommensurate peaks merge into one due to their finite width, resulting in a broad commensurate peak elongated in the transverse direction. Similar measurements of strongly overdoped samples are therefore necessary to confirm this scenario and the emerging similarity to the 11-compounds.\enlargethispage{3pt}

\begin{figure}[t]
\includegraphics[width=1\columnwidth]{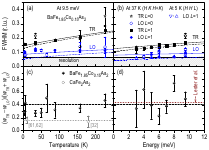}
\caption{(a)~Longitudinal (LO) and transverse (TR) widths of the commensurate peaks around $Q_{\rm AFM}$ ($L=1$) and $Q_\parallel$ ($L=0$) at 9.5\,meV versus temperature. (b)~The same widths versus energy transfer at low temperatures. Solid lines are results of a global fit to the data in both panels (see text) using Eq.\,(\ref{Eq:WidthFitting}). Resolution-corrected dependencies are shown by dashed lines. (c)~The resolution-corrected anisotropy ratio $A=(w_{\rm TR}-w_{\rm LO})/(w_{\rm TR}+w_{\rm LO})$ as a function of temperature, compared to the respective values for the magnetically ordered \cite{ZhaoAdroja09, DialloAntropov09} and paramagnetic \cite{DialloPratt10} states of CaFe$_2$As$_2$. (d)~The same ratio as a function of energy transfer. The dashed line gives the anisotropy from the global fit. The dotted line is derived from the high-energy dispersion reported for BaFe$_{1.87}$Co$_{0.13}$As$_2$ in Ref.\,\onlinecite{LesterChu10}.}
\label{Fig:Anisotropy}
\vspace{-10pt}
\end{figure}

\begin{figure*}\vspace{-1em}
\includegraphics[width=\textwidth]{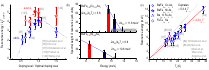}
\caption{(a) Doping dependence of \Er\ at odd and even $L$ in BFNA and BFCA studied here (full symbols) and in previous works (empty symbols), as referred to in (a) and (c). The blue line follows the average $T_{\rm c}$, rescaled to $4.3$ at its optimum \cite{ChuAnalytis09, LesterChu09, CanfieldBudko09, NiThaler10}. The red line is a guide to the eye. (b) Linear extrapolation of the resonance intensities to the energy axis, as compared to the onset of particle-hole continuum. The hatched region covers the range of directly measured $2\Delta$ values for the larger gap in nearly optimally doped BFCA, estimated by various experimental techniques \cite{TerashimaSekiba09, SamuelyPribulova09, YinZech09, HeumenHuang10, KimRossle10, HardyWolf10, HardyBurger10}. (c) Resonance energy versus $T_{\rm c,opt}$ for different Fe-based superconductors. For the compounds with dispersing resonance, only \Er\ at odd $L$ is shown.}
\vspace{-10pt}
\label{Fig:Scaling}
\end{figure*}

In order to quantify the observed in-plane anisotropy and compare it with previous experiments, in Fig.\,\ref{Fig:Anisotropy} we plot the temperature and energy dependence of the measured full width at half maximum (FWHM) of the commensurate inelastic peak along the longitudinal (LO) and transverse (TR) directions for $L=0$ and $L=1$. In the longitudinal direction, the resolution-corrected width of the peaks $w_{\rm LO}$ (dashed line) was already quantified for the same sample by a fit to the Moriya formula \cite{Moriya85} in Ref.\,\onlinecite{InosovPark10}. To extract the anisotropy ratio $A=(w_{\rm TR}-w_{\rm LO})/(w_{\rm TR}+w_{\rm LO})$, we have fitted the experimentally measured FWHM of the peaks in the longitudinal ($W_{\rm LO}$) and transverse ($W_{\rm TR}$) directions (solid lines in Fig.\,\ref{Fig:Anisotropy}) using the following equations:

\noindent\begin{equation}\label{Eq:WidthFitting}
\begin{split}
W_{\rm LO}(\omega,\,T)&= \sqrt{w_{\rm LO}^2(\omega,\,T)+R^2};\\
W_{\rm TR}(\omega,\,T)&= \sqrt{\Biggl[\frac{1+A}{1-A}\,w_{\rm LO}(\omega,\,T)\Biggr]^2\kern-3pt+R^2}.
\end{split}
\end{equation}
The fitted value of the effective resolution, $R=0.066\pm0.004$~r.\,l.u., was used to perform resolution correction of the experimental data and calculate the anisotropy ratio that is presented in panels (c) and (d). By setting the reso\-lution to a constant, we relied on the fact that the calculated instrumental resolution is nearly isotropic and does not vary within our region of interest by more than $\sim$\,10\%. The effective momentum-space resolution resulting from our fit (hatched region in Fig.\,\ref{Fig:Anisotropy}) is somewhat lower than the calculated instrumental resolution ($R_{\rm min}\approx0.04$~r.\,l.u.). The difference may indicate a finite-size limit on the fluctuating domains imposed by the random distribution of dopant atoms and/or a slight inhomogeneous broadening due to variations of the doping level across the sample. With this reasonable assumption, the entire data set can be described by a single, temperature- and energy-independent anisotropy parameter. A similar fit based on the instrumental resolution alone (without finite-size or inhomogeneous broadening) would yield an anisotropy parameter that increases with temperature, which would be highly unusual.

The anisotropy ratio $A=0.41\pm0.02$ that results from the global fit to our data is shown in Fig.\,\ref{Fig:Anisotropy}~(c) and (d) by the dashed line. This value corresponds to the aspect ratio $w_{\rm TR}/w_{\rm LO}\,=\,2.4\pm0.1$, which is nearly a factor of 2 larger than the respective ratio of spin wave velocities ($\sim$1.4) in the undoped CaFe$_2$As$_2$, according to Refs.~ \onlinecite{ZhaoAdroja09} and~ \onlinecite{DialloAntropov09}. The dotted line in Fig.\,\ref{Fig:Anisotropy}\,(c) shows that the anisotropy ratio remains nearly constant across the SDW transition, as estimated from the paramagnetic-state data measured at $T=180$\,K by Diallo \textit{et al} \cite{DialloPratt10}. On the other hand, the anisotropy ratio of 0.44 extracted from the high-energy TOF data on a similarly doped BaFe$_{1.87}$Co$_{0.13}$As$_2$ compound \cite{LesterChu10} [dotted line in Fig.\,\ref{Fig:Anisotropy}\,(d)] perfectly coincides with our value. This agreement confirms the energy independence of the anisotropy and indicates that the difference in the peak widths originates mainly from two unresolved incommensurate peaks, in agreement with our DFT calculations, rather than from an anisotropic broadening caused by the finite correlation lengths of the spin excitations \cite{LiBroholm10, DialloPratt10}. Despite the present lack of $\mathbf{Q}$-resolved INS data on hole-doped compounds, the results of our susceptibility calculations from section \ref{SubSec:DFT} allow us to predict that the anisotropy of the spin-excitation spectrum should vanish and subsequently switch to the longitudinal orientation as the system is doped with more holes.

\vspace{-5pt}\subsection{Doping dependence of the resonance}\label{SubSec:DopDep}\vspace{-5pt}

In order to investigate the doping dependence of the resonance and its $L$-modulation, we summarize in Fig.\,\ref{Fig:Scaling}\,(a) our results together with other studies of electron-doped \BFA\ (no momentum-resolved data for hole doping \cite{FootnoteSeparation} are available so far) \cite{LumsdenChristianson09, ChiSchneidewind09, ChristiansonLumsden09}. To put the \Er\ values from different compounds and doping levels on the same scale, we divided \Er\ by the optimal $k_{\rm B}T_{\rm c,opt}$ and normalized the doping level by the optimal doping level, respectively. While \Erodd\ values (blue symbols) in $\mathbf{Q}=(\frac{1}{\text{\raisebox{0.8pt}{2}}} \frac{1}{\text{\raisebox{0.8pt}{2}}} L)$ fall onto the blue dotted line which follows the average $T_{\rm c}$ in the phase diagrams from Refs.\,\onlinecite{ChuAnalytis09, LesterChu09, CanfieldBudko09, NiThaler10}, \Ereven\ values (red symbols) do not follow $T_{\rm c}$, but rather stay at higher energies than \Erodd\ in the underdoped region, in agreement with a similar recent study \cite{WangLuo10}. As a consequence, the difference between \Erodd\ and \Ereven\ increases with underdoping (as can also be seen in Fig.\,\ref{Fig:Normal-vs-SC}).\enlargethispage{0.5em}

The integrated intensity of the resonance is influenced, in particular, by its proximity to the particle-hole continuum with an onset at $2\Delta$. As a consistency check, we therefore plot in Fig.\,\ref{Fig:Scaling}\,(b) the $Q$- and $\omega$-integrated intensities of the resonance at odd and even $L$ versus its energy. Since in an RPA description the spectral weight of the resonant mode is roughly proportional to its excitonic binding energy \cite{MillisMonien96, PailhesSidis03, PailhesUlrich06}, under the assumption of $L$-independent onset of the particle-hole continuum, a linear extrapolation of the two intensities onto the energy axis gives us a rough lower estimate of $2\Delta$\,---\,the point where the resonance intensity is fully suppressed by particle-hole scattering (for similar analysis in cuprates, see Ref.\,\onlinecite{PailhesUlrich06}). For the Co-doped compound, such an extrapolation results in $2\Delta_{\rm BFCA}\approx11.8$\,meV, which indeed falls in the middle of the range of values reported from direct measurements \cite{TerashimaSekiba09, SamuelyPribulova09, YinZech09, HeumenHuang10, KimRossle10, HardyWolf10, HardyBurger10} (hatched region). Since SC gap measurements for the Ni-doped compound are scarce, we resort to calculating the coupling constant $2\Delta/k_{\rm B}T_{\rm c}=6.8$ that results from the extrapolated gap of $2\Delta_{\rm BFNA}\approx10.6$\,meV. On the one hand, it agrees with the universal value of $7\,\pm2$ that was reported for the larger gap in various two-gap ferropnictides \cite{EvtushinskyInosov09NJP} and coincides with that of 6.8 (or 6.6) derived from combined ARPES and muon-spin rotation ($\mu$SR) \cite{EvtushinskyInosov09, KhasanovEvtushinsky09} and specific-heat \cite{PopovichBoris10} measurements on \BKFA, respectively. On the other hand, it exceeds the maximum coupling constant of $2\Delta/k_{\rm B}T_{\rm c}\approx5.0$ that was recently inferred \cite{HardyWolf10, HardyBurger10} from specific-heat measurements on BFCA. The non-linear dependence of the larger gap on $T_{\rm c}$, reported in Ref.\,\onlinecite{HardyBurger10}, would result in a much lower estimate for $2\Delta_{\rm BFNA}\approx6.9$\,meV in the Ni-doped sample, under the assumption that this dependence is universal among 122-compounds. Such low value would imply a considerable overlap of the resonance peak with the particle-hole continuum, which could explain its broad width in energy.

The successful application of the simple scaling relation with $L$-independent particle-hole continuum indicates that the distance between the resonance and the continuum $2\Delta - \omega_{\rm res}$ is $L$-dependent, as otherwise the agreement with directly measured gap values would be coincidental. In other words, the $L$-dependence of the resonance energy and intensity alone does not necessarily imply a $k_z$-dependent energy gap, as suggested previously \cite{ChiSchneidewind09}, but more likely is a natural consequence of the normal-state intensity modulation. While a SC order parameter that differs at odd and even $L$ values is conceivable and was even supported by experimental evidence \cite{ZhangYang10, WangQian10, EvtushinskyCommunication}, it can only result from the normal-state properties of the ``pairing glue'', and thus does not appear to be the primary reason for the dispersing resonant mode.

Finally, in Fig.\,\ref{Fig:Scaling}\,(c) we combine our data with all the previously reported data \cite{ChristiansonGoremychkin08, LumsdenChristianson09, ChiSchneidewind09, ChristiansonLumsden09, QiuBao09, WenXu10, MookLumsden10, MookLumsden10a, ArgyriouHiess10} to show how the resonance energy \Er\ at odd $L$ depends on $T_{\rm c}$. Filled symbols are extracted from the present work, whereas empty symbols are from the references indicated in the plot. We see that the resonance energy scales linearly with $T_{\rm c}$ from the lowest to the highest critical temperatures with a universal ratio of $\Erodd/k_{\rm B}T_{\rm c} \approx 4.3$ for all hole- and electron-doped Fe-based superconductors investigated so far. It is interesting that such simple linear scaling apparently breaks down at even $L$ values \cite{WangLuo10}, as also seen in Fig.\,\ref{Fig:Scaling}\,(a).

\vspace{-5pt}\section{Discussion and conclusions}

\vspace{-5pt}\subsection{Normal state}\vspace{-5pt}

As we have demonstrated, the elliptical shape of the spin excitations within the $L=\text{const}$ planes shows no measurable $L$-dependence (apart from an intensity modulation) and is insensitive to the SC transition. Therefore, the origins of this anisotropy are to be found in the properties of the normal (paramagnetic) state. An anisotropic spin correlation length that is larger in the direction parallel to the AFM propagation vector than in the transverse (ferromagnetic) direction has been proposed as a possible explanation \cite{LiBroholm10, DialloPratt10}. Although such description is successful in the low-energy region, where the two spin wave branches are not resolved, it clearly fails to describe the anisotropic spin wave velocities that become evident at higher energies in the paramagnetic state \cite{LesterChu10, LiBroholm10}, mimicking the behavior of the parent compounds \cite{DialloAntropov09}. This implies that the larger momentum width of the spectrum in the transverse direction is more likely to be a result either of two unresolved spin-wave branches that are less steep than the longitudinal ones, or of the incommensurability of the nesting peaks at $\omega=0$. The results of our DFT calculations support the incommensurate nesting scenario, similar to that inferred earlier from nuclear-magnetic-resonance measurements \cite{LaplaceBobroff09} and to the one proposed for the iron chalcogenides \cite{ArgyriouHiess10, LeeXu10, LiZhang10, LiuHu10}. In such a case, the anisotropy results from FS nesting, and not from an ``electronic liquid-crystal state'' that arises spontaneously from electron-electron interactions \cite{FangYao08, LesterChu10, ChuangAllan10, KnolleEremin10, ChuAnalytis10}. The latter state has been invoked for the cuprates based in part on the strong temperature dependence of the in-plane anisotropy of the spin excitations \cite{HinkovHaug08}, which is not observed in the 122-iron-arsenide system (Fig.\,\ref{Fig:Anisotropy}). If our prediction of the rotated (longitudinally elongated) susceptibility profile in the hole-doped compounds were confirmed experimentally, it would provide additional support for this scenario.

Although the exact causes for the anisotropy are still under debate, it has been argued that remnant magnetism persisting above the AFM transition, in the form of fluctuating magnetic domains \cite{MazinJohannes09} or the above-mentioned ``electronic nematic'' ground state, is involved. However, as we have demonstrated, the spin-fluctuation spectrum possesses considerable anisotropy and does not fully follow the crystallographic symmetry even in the normal (paramagnetic and tetragonal) state, when no electronic nematicity is assumed. Therefore, the observed in-plane anisotropy in the doped compounds does not necessarily imply a symmetry-broken ground state, but has a more trivial structural origin. As the primitive structural unit cell of the 122-compounds contains two Fe atoms, its size is twice larger, as compared to that of the Fe-sublattice. Because the magnetic INS signal originates predominantly from the latter, with no magnetic moment being induced on the As sites \cite{LeeVaknin10, BrownChatterji10}, the symmetry of the spectrum is determined by the unfolded BZ. In the real bct BZ, both the electronic bands and the spin susceptibility are folded, but the matrix elements that are responsible for the intensities of the primary features and their replica (an analog of the dynamic structure factors) are such that no abrupt change in the magnetic spectrum can be seen as long as the folding potential remains sufficiently weak. Similar effects were observed and discussed in relationship to the single-particle spectral function (see, for example, Ref.\,\onlinecite{ZabolotnyyKordyuk09} and references therein), where matrix elements are also famous for shaping some of the spectral features \cite{InosovFink07, InosovSchuster08}.

The periodic modulation of the magnetic spectral weight with $L$ can also be explained by the $L$-dependence of the spin susceptibility observed in our normal-state DFT calculations. Although the variation of the Lindhard function between $L=0$ and $L=1$ is weak in the parent compound, it can possibly be enhanced by the Stoner-like renormalization effects to an amplitude comparable with experimental observations. The maximum of $\mathrm{Re}\,\chi_0(\mathbf{Q},0)$ in the parent compound occurs at $\mathbf{Q}_\text{AFM}=(\frac{1}{\text{\raisebox{0.8pt}{2}}} \frac{1}{\text{\raisebox{0.8pt}{2}}} 1)$ and thus determines the AFM ordering wavevector.

In the SDW state, excitations at even $L$ correspond to zone-boundary magnons which, due to a combination of intra- and interlayer coupling parameters, have a substantial gap of $\sim 80$\,meV \cite{McQueeneyDiallo08, ZhaoYao08, MatanMorinaga09, ZhaoAdroja09, DialloAntropov09}. In contrast, at high doping levels the magnetic response is virtually $L$-independent \cite{LumsdenChristianson09}. Two mechanisms are likely to provide the connection between these two limiting cases: First, when approaching the magnetically ordered state from higher doping levels, the paramagnon mode softens at $\mathbf{Q}_\text{AFM}$, and the in-plane magnetic correlation length increases \cite{BradenSidis02}. As a consequence, one can expect the out-of-plane magnetic correlations to become more efficient in stabilizing the mode and its gapped response at even $L$. Second, when starting from the ordered state, the increasing damping of the mode with doping will progressively redistribute spectral weight towards lower energies, including the gapped region around even $L$ \cite{PrattKreyssig10}. At our intermediate doping levels we thus observe a moderate $L$-modulation in the normal state [Figs.\,\ref{Fig:Normal-vs-SC} and \ref{Fig:Scaling}\,(a)].

Starting from the paramagnetic state, one sees that the similarity of excitation spectra in the magnetically ordered and normal states does not imply that the anisotropy of the SDW state survives above the structural transition in the form of ``spin nematic correlations''. On the contrary, the symmetry-breaking $L$-modulation is present in the tetragonal phase for reasons not related to magnetic ordering, whereas the SDW instability that occurs on top of the paramagnetic state upon cooling or decreasing the doping is predetermined by this modulation, so that the AFM propagation vector coincides with the strongest nesting-driven peak in $\mathrm{Re}\,\chi_0(\mathbf{Q},0)$. An electronic nematic state also appears implausible in view of the temperature independence of the in-plane anisotropy (Fig.\,\ref{Fig:Anisotropy}), which is in sharp contrast to the strongly temperature dependent, order-parameter-like behavior observed in \ybco\ (Ref.\,\onlinecite{HinkovHaug08}). Our conclusions about the nonmagnetic origin of the missing symmetry are additionally supported by the following evidence: (i) experimentally observed enhancement of the anisotropy in the doped compound with respect to a magnetically ordered parent, which agrees with the increased transverse incommensurability seen in the DFT calculations; (ii) temperature-independence of the anisotropy even in the parent compound, including its insensitivity to the presence of static AFM order. Independently of its origins, the symmetry of the normal-state spin-fluctuation spectrum may have important implications for the SC order parameter under the assumption of spin-fluctuation-driven superconductivity. It was argued, for example, that the transverse elongation of the spin-fluctuation profile stabilizes the $s^\pm$ pairing state \cite{ZhangSknepnek10}.

\vspace{-5pt}\subsection{Superconducting state}\label{SubSec:SCstate}\vspace{-5pt}

Now, let us consider the SC state properties and discuss the implications of our results for the SC pairing mechanism. First, we note that while the conventional unit cell contains two FeAs layers, the primitive cell, from which the BZ is constructed, contains only one. Thus, in contrast to cuprates only one resonant mode is expected; the different resonance energies at even and odd $L$ are therefore to be attributed to an $L$-dispersion rather than to a mode splitting, in agreement with a recent report \cite{PrattKreyssig10}. This dispersion signals the non-negligible 3D character of the electronic band structure and its importance for the description of
the SC state. Indeed, there is compelling evidence for such three-dimensionality both from ARPES \cite{VilmercatiFedorov09, LiuKondo09, MalaebYoshida09, BrouetMarsi09, XuHuang10, ZhangYang10, YoshidaNishi10} and band structure calculations \cite{Singh08, WangQian10}.

Since the calculations we have presented here are limited to the normal state, our discussion of the magnetic spectrum in the SC state has to remain on a qualitative level. However, in view of the normal-state $L$-modulation, already a minimal model like RPA is expected to capture the $L$-dispersion of \Er: By virtue of the resonance condition, namely the vanishing denominator in $\chi_0(\mathbf{Q},\omega)/[1-I(\mathbf{Q}) \chi_0(\mathbf{Q},\omega)]$, the modulation is carried over into the SC state. Here both the bare susceptibility $\chi_0$ and the interaction $I$ can depend on $L$. We keep $I$ deliberately general\,---\,there is no need to refer, for instance, to a $t$-$J$ model \cite{ChiSchneidewind09, PrattKreyssig10}, whose applicability to the iron arsenides is being disputed.

\begin{figure}[t]
\includegraphics[width=1\columnwidth]{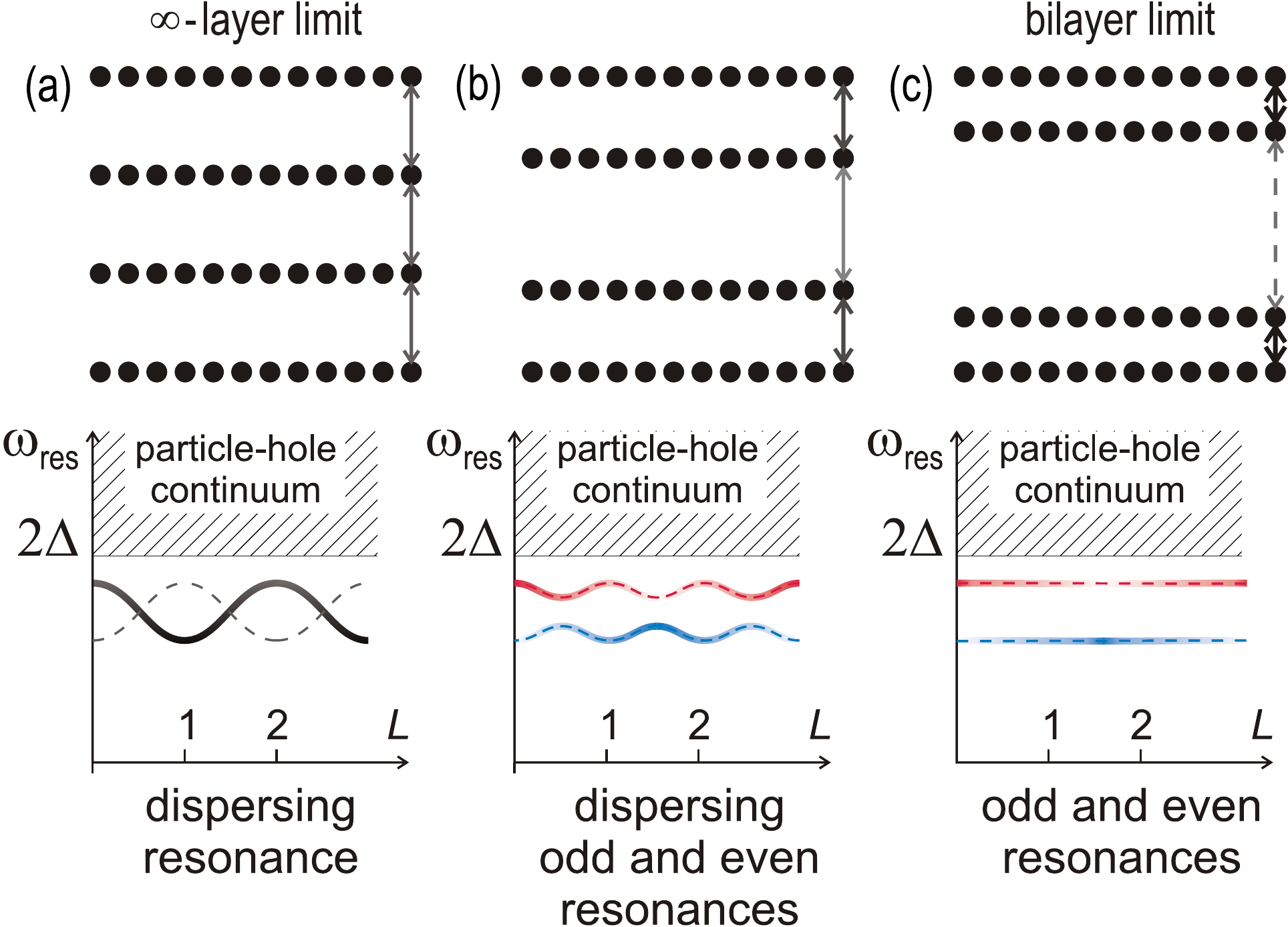}
\caption{Illustration of the evolution from an $\infty$-layer system like BFCA to a bilayer system like YBCO in terms of inter- and intra-bilayer distances and effective interactions. For the equidistant limit (left), a single dispersing resonant mode is observed, whose intensity modulation (shown here by the brightness of the curve) is mainly governed by the closeness to the particle-hole continuum with an onset at $2\Delta$. The dashed line depicts the replica that gains intensity only after the equivalency of the layers is broken (middle panel). For alternating interlayer coupling, the resonance splits into odd and even modes, which become non-dispersive for the case of \ybco\ with nearly independent bilayers (right).}
\label{Fig:Analogy}
\vspace{-12pt}
\end{figure}

We now put our considerations into a broader context by comparing our results to the resonant phenomena in \ybco\ (Fig.\,\ref{Fig:Analogy}). The latter consists of nearly independent CuO$_2$ bilayers and exhibits manifestly 2D electronic structure and SC gap. One observes two distinct, non-degenerate resonances due to the difference in both the bare susceptibility $\chi_0$ and the interaction $I$ between the even and odd channels \cite{EreminMorr07, ZhouWang07, InosovBorisenko07}, which can be ultimately tracked back to the contrast between the intra- and interbilayer hopping and interaction terms. On the contrary, in our iron-arsenide samples, this contrast vanishes and we observe a single resonance, which in addition disperses for the reasons described above. Thus, both systems represent different limiting cases of a more general model with coupled bilayers and possibly 3D electronic structure, Fig.\,\ref{Fig:Analogy}\,(b), where we expect two resonant modes which both disperse and exhibit an intensity modulation along $L$, depending on the effective coupling.

The similarity between the doping dependence of the out-of-plane dispersion bandwidth in the 122-family of iron arsenides [Fig.\,\ref{Fig:Scaling}\,(a)] and the even-odd resonant-mode splitting in bilayer cuprates \cite{PailhesUlrich06, CapognaFauque07} supports our juxtaposition of the two systems: In both cases, the even-odd difference increases when moving towards the magnetic quantum critical point. Whereas the vanishing difference in Fig.\,\ref{Fig:Scaling}\,(a) around optimal $T_\text{c}$ suggests that it is determined by the proximity to the magnetic instability, emphasizing the importance of the out-of-plane magnetic coupling in the arsenides, recent measurements suggest a persistent even-odd difference even beyond optimal doping level \cite{WangLuo10}, indicating that it rather scales with $T_{\rm c}$. More detailed experimental and theoretical work is necessary to settle this point.

Next, we address the implications of the linear relationship between \Er\ and $T_{\rm c}$ with $\Erodd/k_{\rm B}T_{\rm c}\approx4.3$, Fig.\,\ref{Fig:Scaling}\,(c). First, the lower value of this ratio, as compared to that of 5.3 for cuprates, supports the notion of a weaker SC pairing in Fe-based superconductors \cite{InosovPark10}. Second, the validity of the linear relationship for all Fe-based superconductors hitherto studied, independent of the doping carrier type and over the entire studied doping range \cite{ChristiansonGoremychkin08, LumsdenChristianson09, ChiSchneidewind09, ChristiansonLumsden09, QiuBao09, WenXu10, MookLumsden10, MookLumsden10a, ArgyriouHiess10, InosovPark10}, suggests that models that attribute the resonant mode to an excitonic bound state within the SC gap may be more straightforwardly applicable to Fe-based superconductors than they are to the cuprates. Whereas in cuprates deviations from the linear relationship accompany the increasingly anomalous physical properties at underdoping \cite{MookDai02, HinkovBourges07}, the resonance in Fe-based superconductors is remarkably insensitive to the proximate magnetic state and even coexists with it at very low doping \cite{PrattKreyssig10}.

Finally, we remark that while in a recent report a large $L$-dispersion bandwidth was related to the presence of long-range magnetic order or pronounced spin correlations \cite{PrattKreyssig10}, here we observe appreciable bandwidth (only $\sim35\%$ less than in Ref.\,\onlinecite{PrattKreyssig10}) in a paramagnetic compound, which we associate with the normal-state intensity modulation that can be qualitatively reproduced even in the Lindhard function calculated for the nonmagnetic ground state.

\vspace{-10pt}\section*{Acknowledgements}\vspace{-5pt}

We are grateful to O.~K.~Andersen, L.~Boeri, A.~Boris, A.~Charnukha, A.~V.~Chubukov, D.\,V.~Efremov, I.~Eremin, D.~V.~Evtushinsky, G.~Jackeli, D.~Johnston, I.~I.~Mazin, D.~Reznik, and H.~Yamase for helpful discussions and to R.~A.~Mole for the technical support during INS experiments. This work has been supported, in part, by the DFG within the Schwerpunktprogramm 1458, under Grant \mbox{No.\,BO3537/1-1}.

\bibliographystyle{my-apsrev}\bibliography{Nematicity}

\end{document}